\documentclass[twocolumn,twocolappendix]{aastex701}
\usepackage{graphicx} % Required for inserting images
\usepackage{gensymb}
\usepackage{enumitem}
\defcitealias{idmage}{Paper I}

\newcommand{\hi}{H\,{\sc i}}

\newcommand{\kms}{km~s$^{-1}$}

\newcommand{\msun}{M_\odot}

\newcommand{\candsNGC}{33}

\newcommand{\totcands}{166}

\newcommand{\highcands}{66}
\newcommand{\fullcands}{100}
\newcommand{\photcut}{59}
\newcommand{\background}{72}
\newcommand{\followedup}{78}

\newcommand{\gbtobs}{82}
\newcommand{\gbtdetect}{19}

\newcommand{\gbtnodetect}{63}
\newcommand{\hisats}{18}

\newcommand{\LMCLFerup}{3.1$\pm$1.2}
\newcommand{\SMCLFerup}{1.7$\pm$0.6}

\newcommand{\LMCLFold}{4.0$\pm$1.4}
\newcommand{\SMCLFold}{2.1$\pm$0.6}

\newcommand{\lowlimsm}{1.2$\pm$0.4}
\newcommand{\lowlimlm}{1.7$\pm$0.7}

\newcommand{\lowlimsmup}{1.0$\pm$0.3}
\newcommand{\lowlimlmup}{1.7$\pm$0.7}

\newcommand{\UA}{\affiliation{Steward Observatory, University of Arizona, 933 North Cherry Avenue, Tucson, AZ 85721-0065, USA}}

\newcommand{\Dart}{\affiliation{Department of Physics and Astronomy, Dartmouth College, 6127 Wilder Laboratory, Hanover, NH 03755, USA}}
	
\graphicspath{{figures/}}

\begin{document}

\title{ID-MAGE II: The Star-Forming Satellites of Low-Mass Hosts}

%\correspondingauthor{Laura Congreve Hunter}
%\email{laura.c.hunter@dartmouth.edu}
\author[0000-0001-5368-3632]{Laura Congreve Hunter}
\email{laurachunter@gmail.com}
\Dart

\author[0000-0001-9649-4815]{Bur\c{c}in Mutlu-Pakdil}
\email{Burcin.Mutlu-Pakdil@dartmouth.edu}
%\affil{Department of Physics and Astronomy, Dartmouth College, Hanover, NH 03755, USA}
\Dart

\author[0009-0001-8573-1800]{Michael B. Farnell}
\email{Michael.B.Farnell.III.26@dartmouth.edu}
\Dart

\author[0000-0003-4102-380X]{David~J. Sand}
\email{dave.j.sand@gmail.com}
\UA

\author[0000-0002-0956-7949]{Kristine Spekkens}
\email{kristine.spekkens@queensu.ca}
\affiliation{Department of Physics, Engineering Physics and Astronomy, Queen’s University, Kingston, ON K7L 3N6, Canada}

\author[0000-0001-8354-7279]{Paul Bennet}
\email{pnbennet@gmail.com}
\affiliation{Space Telescope Science Institute, 3700 San Martin Drive, Baltimore, MD 21218, USA}

\author[0009-0007-9488-7050]{Sasha N. Campana}
\email{sncampana@aol.com}
\Dart

\author[0000-0002-3936-9628]{Jeffrey L. Carlin}
\email{jeffreylcarlin@gmail.com}
\affil{AURA/Rubin Observatory, 950 North Cherry Avenue, Tucson, AZ 85719, USA} 

\author[0000-0002-1763-4128]{Denija Crnojevi\'{c}}
\email{dcrnojevic@ut.edu}
\affil{Department of Physics \& Astronomy, University of Tampa, 401 West Kennedy Boulevard, Tampa, FL 33606, USA}

\author[0000-0001-9775-9029]{Amandine Doliva-Dolinsky}
%\Dart
\email{amandinedolinsky@gmail.com}
\affil{Department of Physics, University of Surrey, Guildford, Surrey GU2 7XH, UK}
%\affil{Department of Physics \& Astronomy, University of Tampa, 401 West Kennedy Boulevard, Tampa, FL 33606, USA}

\author[0009-0004-9516-9593]{Emmanuel Durodola}
\email{emmanuel.a.durodola.gr@dartmouth.edu}
\Dart

%\author[0000-0001-8245-779X]{Catherine Fielder}
%\email{fielder.catherine@gmail.com}
%\UA

\author[0000-0002-5434-4904]{Michael G. Jones}
\email{mike.jones.astro@gmail.com}
\affiliation{IPAC, Mail Code 100-22, Caltech, 1200 E. California Blvd., Pasadena, CA 91125, USA}
\UA

\author[0000-0002-7013-4392]{Donghyeon J. Khim}
\email{galaxydiver@arizona.edu}
\UA

%\author[0000-0001-8855-3635]{Ananthan Karunakaran}
%\email{karunakaranananthan@gmail.com}
%\affiliation{Department of Astronomy \& Astrophysics, University of Toronto, Toronto, ON M5S 3H4, Canada}

\author[0009-0008-3389-9848]{Laurella Marin}
\email{emarin4@uw.edu}
\affiliation{Department of Astronomy, University of Washington, Physics-Astronomy Bldg, Seattle, WA}

\author[0009-0004-9126-8260]{Ricardo J. Mendez}
\email{ricardo.j.mendez.26@dartmouth.edu}
\Dart

\author[0000-0002-8217-5626]{Deepthi S. Prabhu}
\email{deepthisprabhu@gmail.com}
\UA

\author[0000-0002-5177-727X]{Dennis Zaritsky}
\email{dennis.zaritsky@gmail.com}
\UA

\begin{abstract}
    We present the first results of the ongoing Identifying Dwarfs of \textbf{Magellanic Cloud} Analog GalaxiEs (ID-MAGE) survey. This includes new velocities of \followedup\ satellite candidates identified around 35~nearby LMC- and SMC-mass hosts (D$=$4$-$10~Mpc). Using new Green Bank Telescope \hi\ observations, new optical long-slit spectroscopy from SALT and MDM~2.4m, and Dark Energy Spectroscopic Instrument DR1, we find six are velocity-confirmed satellite galaxies ($6.5\times10^5\leq M_\star/M_\odot\leq1.5\times10^7$) and \background\ are background galaxies. Our results underscore the ability of spectroscopic follow-up to effectively separate satellites from background galaxies. Including the six newly velocity-confirmed satellites and the 19 satellites confirmed in the literature, the ID-MAGE hosts have a combined total of 25 confirmed satellites. This confirmed satellite sample includes the complete satellite populations of three hosts (UGC~04426: zero satellites, UGC~08201: zero satellites, NGC~3432: four satellites), which we compare to simulations and known satellite systems from previous systematic searches around low-mass hosts. Our confirmed satellite sample is nearly complete for the most massive satellites (M$_\star > 10^7~\msun$). We find these satellites have a quenched fraction of 10--25\%, placing them between the $<$5\% quenched fraction of isolated galaxies and the 40--70\% quenched fraction of MW-analog satellites in the same stellar mass range ($10^7~M_\odot < $ M$_\star < 10^8~\msun$). This demonstrates the impact that low-mass galaxies have on the evolution of their satellites. 
\end{abstract}

\section{Introduction}

Satellite galaxy populations are among the most sensitive astrophysical probes of the nature of dark matter and galaxy evolution. They have produced some of the tightest astrophysical constraints on dark matter \citep[e.g.,][]{Nadler21, Nadler2021b}, making them a key, testable feature of $\Lambda$ Cold Dark Matter ($\Lambda$CDM) models.  Due to their low masses, satellites are also sensitive probes of the physics of galaxy formation \citep[e.g.,][]{Agertz2020,Kim2024, Rey2025}.

Our current understanding of how hosts impact the evolution of their satellite galaxies is primarily based on satellites of MW-mass hosts \citep[e.g.,][]{Chiboucas09, Chiboucas2013, Martin2013, Sand14, Crnojevic16, toloba16, SAGAI, Smercina18, Bennet19, carlsten19a, Crnojevic2019, Muller2019, Bennet20, BMP22, ADD23, BMP24, SAGAIII, Danieli25, Tan2025, ADD25_rev}. Recent surveys --- e.g., Exploration of Local VolumE Satellites Survey (ELVES; \citealt{ELVES}) and Satellites Around Galactic Analogs (SAGA; \citealt{SAGAI,SAGAIII}) --- have proven the scientific value of a statistical sample of satellite galaxies. The results of these searches have been used to constrain structure formation within the $\Lambda$CDM model, and to assess the environmental impact on the satellites’ star formation histories (SFH), \hi-gas richness, and morphology  \citep[e.g.,][]{Carlsten2021b, Karaunkaran22, Greene23, Danieli23, Karunakaran23, Jones24a, SAGAIV, SAGAV, zhu2025}.  

Satellites around MW-mass hosts have evolved in largely similar environments, experiencing similar tidal fields and undergoing ram-pressure stripping due to their hosts' hot gaseous coronae \citep{Wetzel2015}.  Uncertainties in dwarf galaxy formation and dark matter models are degenerate with these environmental effects.  There is thus a danger of ``over-tailoring'' these models to fit satellites of MW-mass hosts alone. Due to weaker tidal forces and ram-pressure stripping, satellites of dwarf galaxies may experience a different degree of environmental influence \citep{spekkens14}. Satellites of dwarf galaxies may be quenched through starvation instead of quenching via ram-pressure stripping \citep{garling24}. Starvation is a process by which galaxies slowly deplete their \hi\ reserves and are prevented from accreting pristine cold gas to replenish them \citep{Larson1980, Peng15, Cortese21}. While LMC-mass hosts may quench the star formation activity of their satellites, the SFH and quenching timescales of satellites around LMC-mass hosts are likely more diverse than those of satellites of MW-mass galaxies \citep{Jahn19, Jahn22}. As such, observational studies of the star formation properties and gas kinematics of satellites around low-mass galaxies will provide key insights into the satellite quenching process.

Recent observational studies have made great strides in identifying satellites and satellite candidates around low-mass galaxies \citep{Sand15, sand24, Rich12, Carlin21, Carlin24, Davis2021, Davis2024, mcnanna24, ADD3109,  Medoff25, Liddo161, ELVES-Dwarf, stierwalt25}. Through resolved star searches of nearby hosts, such as the Magellanic Analog Dwarf Companions and Stellar Halos survey (MADCASH; \citealt{MADCASH}), the DEEP component of DECam Local Volume Exploration Survey (DELVE-DEEP; \citealt{DELVE}), and unresolved light searches of hosts, including the Large Binocular Telescope Satellites of Nearby Galaxies Survey (LBT-SONG; \citealt{LBT-SONG,lbtsong2}), LBT Imaging of Galactic Halos and Tidal Structures survey (LIGHTS; \citealt{LIGHTS2024}), Identifying Dwarfs of MC Analog GalaxiEs (ID-MAGE; \citealt{idmage}), and ELVES-Dwarf \citep{ELVES-Dwarf}, the community is working toward a statistical census of satellites of dwarf galaxies. Satellites are being found around hosts in a variety of environments --- from very isolated hosts to members of dwarf galaxy groups --- and are beginning to provide key insights into how environment drives evolution. 

ID-MAGE is an ongoing survey designed to identify and characterize the satellites of 35 low-mass $10^{7.9} M_\odot < M_\star < 10^{9.7} \msun$) host galaxies (D$=$4--10~Mpc) in a variety of environments through an integrated light search \citep{idmage}. In \citet[hereafter \citetalias{idmage}]{idmage}, we identified a list of 355  candidate satellites, including 16 known satellites and 339 new candidate satellites around low-mass hosts. ID-MAGE nearly tripled the number of LMC- and SMC-mass hosts surveyed for satellites with well-characterized detection limits. However, to use the candidate satellite populations as a test for the theoretical predictions of $\Lambda$CDM \citep[e.g.,][]{Dooley17, Nadler2022, Jahn22}, follow-up observations to remove background contaminants are required. % a pure satellite sample is required. 

Previous satellite searches have demonstrated the importance of confirming satellite candidates, as background contamination rates can be as high as $\sim$80\% in searches around MW-mass galaxies (e.g., \citealt{Zaritsky93,Bennet19,Bennet20}).  For many of our candidates, we are conducting a deep imaging campaign to measure distances using surface brightness fluctuations (SBF). SBF uses deep ground-based photometry to accurately measure the distances of red and featureless dwarf galaxies well beyond the Local Group \citep{Fast-SBF}. However, SBF is not ideal for gas-rich, star-forming systems because stars form in clusters, which breaks the key assumption of SBF: that the stars are Poisson distributed \citep{Greco21}. Therefore, for our bluer candidates with patchy morphology, we are simultaneously undertaking a ground-based spectroscopic follow-up campaign to obtain velocity measurements. The SAGA survey demonstrated that the relative velocities between hosts and satellite candidates provide an effective method for confirming star-forming satellites \citep{SAGAI,SAGAIII}. Large velocity offsets are particularly useful for identifying background galaxies, as a candidate with a sufficiently large relative velocity ($\Delta v_{r} \gtrsim$ 300~\kms\ for MW-mass hosts) is unlikely to be gravitationally bound to the host. %However, velocity measurements are an efficient method to separate background galaxies from probable satellites. 

This paper presents our first set of follow-up observations of ID-MAGE satellite candidates, completing follow-up for three hosts and providing a nearly complete sample of the most massive satellites $M_\star\geq10^7$) for all hosts. We compare the satellite systems of our hosts with model predictions, and examine the star formation and gas properties of the confirmed satellites in the context of dwarf galaxies in other environments. n Section~\ref{summary}, we summarize the ID-MAGE survey and candidate sample presented in \citetalias{idmage}. In Section~\ref{observations}, we describe the velocity measurements used to separate satellite and background galaxies.  In Section~\ref{discussion}, we compare our hosts with completed spectroscopic follow-up observations to simulations and other satellite searches, and discuss the quenched fraction and \hi\ properties of the most massive satellites in the sample.  Finally, we summarize our key results in Section~\ref{conclusions}.

\section{Summary of the ID-MAGE Survey}\label{summary}

\begin{table*} 
    \small
    \caption{ID-MAGE Host Properties} \label{table:hosts}
        \begin{tabular}{l c r c c c r} 
        \hline
        \hline 
        Galaxy & Dist & Vel & Confirmed & High-Likelihood  & Extended Sample & Velocity \\ 
         & Mpc & \kms\ & Satellites & Candidates & Candidates & Cite \\
         (1) & (2) & (3) & (4) & (5) & (6) & (7) \\
        \hline 
        \multicolumn{7}{c}{LMC-Mass Hosts}\\ 
        \hline
        NGC~4449 & 4.16$\pm$0.02 & 207$\pm$4 & 1 & 2 & 7 & STMM \\
        NGC~4244 & 4.20$\pm$0.14 & 245$\pm$1 & 0 & 4 & 3 & SHGK \\
        NGC~4605 & 5.41$\pm$0.05 & 155$\pm$5 & 1 & 1 & 0 &SHGK \\
        NGC~6503 & 6.12$\pm$0.20 & 25$\pm$1 & 1 & 1 & 3 & GMA \\
        \textit{NGC~0672$^*$} & 7.00$\pm$0.26 & 429$\pm$1 & \textit{1} & \textit{6} & \textit{7} & SGH \\
        NGC~0024 & 7.13$\pm$0.10 & 554$\pm$1 & 1 & 1 & 1 & SGHK \\
        \textit{IC~1727$^*$} & 7.29$\pm$0.20 & 345$\pm$1 & \textit{1} & \textit{6} & \textit{7} & GMA \\
        \textbf{NGC~3432} & 8.9$\pm$0.80 & 616$\pm$4 & 4 & 0 & 1 & RC3 \\
        NGC~7090 & 9.29$\pm$0.26 & 847$\pm$3 & 0 & 4 & 1 & HIPASS \\[1ex]
        \hline
        \multicolumn{7}{c}{SMC-Mass Hosts}\\
        \hline 
        NGC~0625 & 3.92$\pm$0.07 & 396$\pm$3 & 0 & 2 & 3 & HIPASS \\
        IC~4182 & 4.24$\pm$0.08 & 321$\pm$4 & 2 & 4 & 3 & RC3 \\
        NGC~4236 & 4.31$\pm$0.08 & 0$\pm$3 & 0 & 1 & 7 & RC3 \\
        ESO245-G05 & 4.46$\pm$0.12 & 391$\pm$2 & 0 & 3 & 2 & HIPASS \\
        NGC~5204 & 4.48$\pm$0.50 & 201$\pm$1 & 1 & 1 & 5 & SGHK \\
        NGC~4395 & 4.65$\pm$0.02 & 319$\pm$4 & 0 & 2 & 2 &SGHK \\
        \textbf{UGC~08201} & 4.72$\pm$0.04 & 31$\pm$1 & 0 & 0 & 1 & HKK \\
        ESO115-G21 & 4.96$\pm$0.05 & 515$\pm$2 & 0 & 2 & 0 & HIPASS\\
        %NGC~3738 & 5.19$\pm$0.05 & 236$\pm$1 & OHB \\
        NGC~0784 & 5.26$\pm$0.02 & 198$\pm$1 & 2 & 3 & 2 & HVG \\
        IC~5052 & 5.37$\pm$0.15 & 584$\pm$3 & 0 & 1 & 5 & HIPASS \\
        NGC~1705 & 5.61$\pm$0.10 & 633$\pm$6 & 0 & 1 & 5 & HIPASS \\
        ESO154-G23 & 5.74$\pm$0.05 & 574$\pm$2 & 0 & 1 & 2  & HIPASS \\
        IC~1959 & 6.07$\pm$0.11 & 639$\pm$1 & 0 & 2 & 2 & PBS \\
        NGC~4707 & 6.38$\pm$0.29 & 469$\pm$1 & 1 & 1 & 3 & UZC \\
        NGC~4455 & 6.46$\pm$0.27 & 644$\pm$1 & 3 & 3 & 3 & SGHK \\
        NGC~5585 & 6.84$\pm$0.31 & 303$\pm$1 & 0 & 3 & 5 & SGHK \\
        UGC~04115 & 7.70$\pm$0.11 & 341$\pm$5 & 0 & 1 & 4 & HIPASS \\
        UGC~03974 & 7.99$\pm$0.07 & 272$\pm$1 & 1 & 1 & 0 & SGHK \\
        NGC~2188 & 8.22$\pm$0.23 & 747$\pm$4 & 0 & 1 & 5 &  HIPASS \\
        UGC~05423 & 8.66$\pm$0.12 & 347$\pm$1 & 2 & 1 & 5 & HKK \\
        ESO364-G29 & 8.81$\pm$0.33 & 787$\pm$3 & 1 & 3 & 1 & HKK01 \\
        IC~4951 & 9.0$\pm$0.6 & 814$\pm$2 & 0 & 3 & 1 & HIPASS \\
        HIPASSJ0607-34 & 9.4$\pm$0.4 & 765$\pm$7 & 0 & 2 & 1 & HIPASS \\
        \textbf{UGC~04426} & 9.62$\pm$0.18 & 396$\pm$1 & 0 & 0 & 3 & SGHK  \\
        ESO486-G21 & 9.7$\pm$1.3 & 832$\pm$1 & 0 & 4 & 6 & TCH \\
        NGC~4861 & 9.71$\pm$0.18 & 835$\pm$1 & 3 & 1 & 1 & vEMK \\[1ex]
        \hline 
    \end{tabular}
    \tablecomments{Column~1: Galaxy Name. Column~2: TRGB distances in Mpc \citep{EDD}. Column~3: Recessional velocity in \kms. Column~4: Number of Confirmed Satellites. Column~5: High Likelihood Candidates to Confirm. Column~6: Extended Sample Candidates to Confirm. Column~7: Citation for velocity GMA: \cite{GHASP}, HIPASS: \cite{HIPASS}, HKK01: \cite{Huchtmeier01}, HKK: \cite{Huchtmeier03}, HVG: \cite{HVG}, %OHB: \cite{Oh15},
    PVB:\cite{PVB}, RC3: \cite{RC3}, SGH: \cite{SGH08}, SHGK: \cite{springbob05}, STMM \cite{STMM}, TCH: \cite{Theureau17}, vEMK: \cite{vEMK}, UZC: \cite{UZC} \\ 
    $\star$ NGC~0672 and IC~1727 are considered the same system due to their proximity.\\
    Hosts in bold are those we have completed follow-up observations for their High-Likelihood candidates.\\
    }
\end{table*}

In \citetalias{idmage}, we analyzed DESI Legacy Survey DR10 imaging around 9 LMC-mass and 27 SMC-mass hosts (see Table~\ref{table:hosts}) out to a projected distance of 150~kpc to identify unresolved satellite candidates. Two hosts (IC~1727 and NGC~0672) are at similar distances (7.3~Mpc and 7.0~Mpc, respectively), similar masses ($M_\star=10^{9.21}~\msun$ and $M_\star=10^{9.18}~\msun$, respectively) and are close on the sky ($\simeq$10\arcmin\ or $\sim$16~kpc) meaning we cannot distinguish one as the primary "host" of the system. As in \citetalias{idmage}, we treat them as a single system with two hosts. We count both as LMC-mass hosts when considering the satellite and satellite candidate population statistics.

Candidates were identified using a modified version of the detection algorithm described in \cite{Bennet17}, and vetted through multiple rounds of visual inspection.  In summary, the algorithm searches wide-field imaging for low-surface-brightness sources after masking bright objects and spatially binning the images. Candidates were visually inspected to remove image artifacts, Galactic cirrus, background galaxy clusters, and other contaminants. Co-authors then rate the likelihood that each candidate is a satellite galaxy. Candidates which pass inspection are sorted into two samples: the High-Likelihood Sample (100\% agreement they are potentially satellites) and the Extended Sample\footnote{referred to as the Full Sample candidates in \citetalias{idmage}} ($\geq66\%$ agreement they are potentially satellites). The full search is complete down to $M_V\lesssim -9$ at the median host distance and $\mu_{0,g}\simeq26$ mag arcsec$^{-2}$.

As part of the candidate vetting in \citetalias{idmage}, we searched the literature for reliable distance measurements (tip of the red giant branch; TRGB or SBF), as well as published velocity measurements. Candidates with $\Delta v_{r} \gtrsim 250$~\kms\ relative to their assumed host, or  {candidates whose distance measurement's 2$\sigma$ error range does not agree with the host's TRGB distance}, were cut from the sample before the final visual-inspection stage.  This literature review found 13 candidates with published velocities and/or distances confirming them as satellites. We also include three additional confirmed satellites from the literature (MAGE~J1052+3635, MAGE~J1246+5136, MAGE~J1259+3456) that were not recovered by our search algorithm because they were masked in the imaging due to their proximity to bright objects. Since \citetalias{idmage}, recently published ELVES-Dwarf SBF distances have confirmed three additional ID-MAGE candidates as satellites \citep{ELVES-Dwarf}.

As discussed in \citetalias{idmage}, NGC~3738\footnote{See Appendix Table~\ref{table:Background_3738} for new velocities of candidates around NGC~3738 we followed up before its removal as a host.} has an unusually high contamination because its field includes NGC~3718, a Seyfert~1 galaxy at 14~Mpc, and a galaxy cluster at 21~Mpc.  We hereafter remove NGC~3738 from the host sample and eliminate \candsNGC\ candidates. The updated sample  {before follow-up observations} is comprised of 35 hosts with 19 confirmed satellites and 303 satellite candidates.

\section{Observations}\label{observations}

\begin{figure}[!tb]
    \centering
    \includegraphics[width=.95\linewidth]{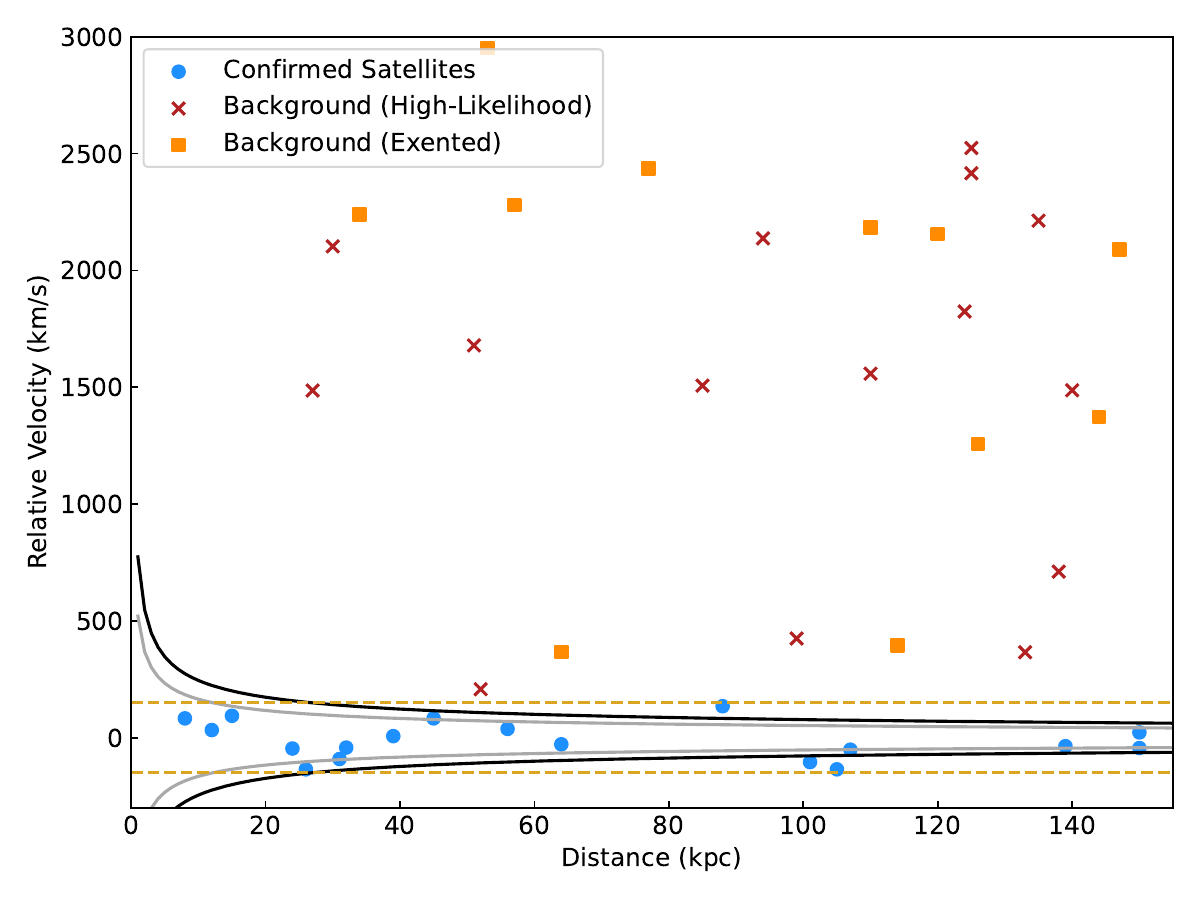}
    \caption{Relative line-of-sight velocity, $v_r$, as a function of projected distance from the host. Blue circles are velocity-confirmed satellites, while red crosses and yellow squares are background galaxies from the High-Likelihood and Extended Samples, respectively. We adopt  {$ \Delta v_{r} \leq$150~\kms\ }(horizontal dashed lines) as the criterion for velocity confirmation. The survey extends to a projected radius of 150~kpc. The black and gray curves show the escape velocities for LMC-mass ($1.8\times10^{11}\msun$) and SMC-mass ($9.5\times10^{10}\msun$) point-mass hosts, respectively, based on the \citet{Brook14} model adopted by \citet{Dooley17}. Of the 72 background galaxies, 46 have $v_r>$3000~\kms\ and fall outside the plotted range.}
    \label{fig:V_v_Sep}
\end{figure}

To follow-up the 303 satellite candidates, we obtained new observations with the Green Bank Telescope (GBT), the South African Large Telescope (SALT), and the MDM Observatory Hiltner~2.4~m telescope. Additionally, after our follow-up campaign began, the DESI DR-1 catalog \citep{desi} was released.  As such, we cross-matched our candidate sample with the DESI catalog to identify new literature velocities.  These first spectroscopic data sets demonstrate the effectiveness of our strategy in removing background galaxies and identifying probable satellites.

To set our velocity range for satellite confirmation, we assume the \citet{Brook14} model virial masses of the LMC and SMC ($1.8\times10^{11}\msun$ and $9.5\times10^{10}\msun$, respectively) as central point masses. In Figure~\ref{fig:V_v_Sep}, we plot their line-of-sight escape velocity versus projected radius in the solid black (LMC-mass) and gray (SMC-mass) lines and select a relative velocity cut of  {$\Delta v_{r} \lesssim 150$~\kms}. No candidates with  {$\Delta v_{r} > 150$~\kms} fall within the escape velocity curves. We therefore classify candidates with  {$\Delta v_{r} \lesssim 150$~\kms} as velocity-confirmed, probable satellites and include them in our Confirmed Satellite Sample.  Our line-of-sight velocity cut corresponds to a three-dimensional velocity of 260~\kms, comparable to the escape velocity at a radius $\simeq$50~kpc for an LMC-mass host.  Whether a velocity-confirmed satellite is in fact bound to its host depends on its orbital configuration and the host mass, but the velocity cut effectively removes background contaminants and provides a mostly pure satellite sample. Throughout this work, we consider satellites with TRGB distances as fully confirmed satellites, and those with velocities or SBF distances as probable satellites because  {both velocity confirmation and SBF distances may yield false positives}. Together, the fully confirmed and probable satellites comprise are our Confirmed Satellite Sample.

\subsection{Green Bank Telescope}\label{green_bank}

We observed \gbtobs\ candidates identified in \citetalias{idmage} with the Robert C. Byrd Green Bank Telescope (GBT) through projects GBT24B-319 and GBT25A-155. The goals of these observations were to both measure the candidates' velocities and determine their \hi\ content.  We selected candidates with no previous \hi\ detections and with evidence of ongoing star formation, as star-forming galaxies are more likely to be gas-rich. We prioritized candidates with GALEX near ultraviolet (NUV) and far ultraviolet (FUV) emission or blue optical colors \textit{g-i ($g-i<0.5$)}. We excluded candidates for which the expected integration time was prohibitively long, based on their stellar mass and host distance. We also cut candidates within $\simeq9'$ of their host or a larger satellite as \hi\ emission from the more massive galaxy could obscure the detection of the candidate.

The planned time-on-source was estimated to reach a 3$\sigma$ detection for a gas fraction of M$_{\text \hi}$/M$\star\geq0.2$, assuming the spectra are smoothed to 30~\kms. The expected \hi\ masses were estimated from the candidate stellar masses in \citetalias{idmage} and the distances of the presumed hosts. Achieving similar \hi\ gas-fraction sensitivity requires short, shallow observations for more massive or nearby candidates and longer, deeper observations for less massive or more distant candidates.  We required a minimum exposure time of 600~s per target to complete one ON/OFF pair and a maximum exposure time of 2.5~hr.

Targets were observed in 600~s ON/OFF pairs, with 300~s on source and 300~s off source. We used the L-band receiver coupled with the Versatile GBT Astronomical Spectrometer (VEGAS) backend in Mode 10, providing a bandwidth of 23.44 MHz and a spectral resolution of 0.7 kHz or 0.15~\kms. The spectral setup was centered at the \hi\ 21~cm line (1.42041~GHz) redshifted to the host's velocity for GBT24B-319 and 500~\kms\ above the host velocity for GBT25A-155. This adjustment allowed us to detect higher velocity background galaxies. In GBT24B-319 the setup covered a recessional velocity range of approximately $-$1000~\kms\ to 3000~\kms\ leaving part of the available bandwidth unused. The data were reduced using the GBTIDL routine \textit{getps}, followed by baseline subtraction and spectral smoothing. We flux calibrated the data using the default L-band noise diode, assuming a typical \textit{getps} flux-scale calibration accuracy of ~20\%. The GBT \hi-fluxes were then multiplied by 1.2 to account for the calibration offset identified in \cite{Goddy2020}. 

For each \hi\ detection, we searched within the 9\arcmin\ GBT beam area for known galaxies at the velocity of the detection. When a match was found, we flagged the spectrum as potentially contaminated. We also cross-matched all GBT detections with the MDM and DESI results to check for velocity consistency. One candidate, MAGE~J1240+6225, has a GBT-derived velocity of 2594$\pm$2~\kms, which does not agree with its DESI DR-1 velocity of 10042~\kms\ \citep{desi}.  {This velocity is outside the frequency range observed by the GBT.} We therefore adopt the DESI velocity and attribute the GBT detection to another galaxy within beam, DESI~J189.8460+62.4098.  

With the GBT, we reliably detect \hi-emission from \gbtdetect\ satellites and candidates.  {For each detected galaxy, we smoothed the spectra from a velocity resolution of 0.15~\kms\ to $~$2.0~\kms\ to improve signal-to-noise. We then fit a gaussian to the emission line to measure the galaxy's systemic velocity.} Two of these detections, MAGE~J0006-2456 and MAGE~J1305+3812, are new velocity-confirmed satellites. One detection, MAGE~J1343+5813 is a previously confirmed satellite without archival \hi\ data, and one, MAGE~J1052+3628, is contaminated by emission from its host, NGC~3432, which lies within $\simeq10'$. The GBT-derived velocity and \hi\ flux for MAGE~J1052+3628 are not reported here because its \hi\ flux cannot be measured reliably.  The remaining 15 detections are background galaxies.  For the \gbtnodetect\ GBT non-detections, we place $5\sigma$ upper-limits on their \hi\ fluxes  {where meaningful and feasible} (see details in Appendix Section~\ref{background}). 

The velocity measurements are listed in Table~\ref{table:velocity_confirmation} for the confirmed satellites and Table~\ref{table:Background_galaxies} for the background galaxies.  The \hi\ fluxes for the confirmed satellites are tabulated in Table~\ref{table:satellite_HI} and included in Table~\ref{table:Background_galaxies} for the background galaxies. Spectra of the newly velocity-confirmed satellites are in Appendix Section~\ref{conf_plots} as Figures~\ref{0006-2456}, \ref{1052+3628}, and \ref{1305+3812}.

\begin{table*}
\caption{New Velocity-Confirmed Satellites}
\hspace*{-2.5cm}
\resizebox{1.1\linewidth}{!}{
\begin{tabular}{l c c r r r r r r l r}
\hline
\hline
Name &  RA & Dec & Tel. & Exp & m$_g$ & $r_e$ & Host & Host Vel. & Sat Vel. & $\Delta v_r$ \\
 & J2000 & J2000 & & Sec & mag & '' & & \kms\ & kms$^{-1}$ & \kms\ \\
(1) & (2) & (3) & (4) & (5) & (6) & (7) & (8) & (9) & (10) & (11)  \\
\hline
MAGE~J0006-2456 & 1.7019 & -24.9446 & GBT & 2400 & 16.0$\pm$0.1 & 7.2$\pm$0.1 & NGC~0024 & 554$\pm$1 &  {573}$\pm$2 & 19 \\
MAGE~J0609-3225 & 92.2913 & -32.4228 & SALT & 2400 & 17.0$\pm$0.1 & 11.1$\pm$0.1 & ESO364-G29 & 787$\pm$3 &  {809$\pm$9} &  {22}\\
MAGE~J1052+3628 & 163.0238 & 36.4773 & DESI & & 17.2$\pm$0.1 & 7.3$\pm$0.1 & NGC~3432 & 616$\pm$4 & 481$\pm$10 & 135 \\
MAGE~J1052+3646 & 163.1554 & 36.7695 & DESI & & 18.4$\pm$0.1 & 6.4$\pm$0.1 & NGC~3432 & 616$\pm$4 & 570$\pm$10 & 46 \\
MAGE~J1052+3646 & 163.1554 & 36.7695 & MDM & 3600 & 18.4$\pm$0.1 & 6.4$\pm$0.1 & NGC~3432 & 616$\pm$4 & 523$\pm$39 & 93 \\
MAGE~J1305+3812 & 196.4781 & 38.2112 & GBT & 7200 & 18.6$\pm$0.1 & 6.7$\pm$0.1 & IC~4182 & 321$\pm$4 &  {405$\pm$1} & 84 \\
MAGE~J1308+3855 & 197.2422 & 38.9185 & DESI & & 18.2$\pm$0.1 & 9.3$\pm$0.1 & IC~4182 & 321$\pm$4 & 270$\pm$10 & 49\\
MAGE~J1308+3855 & 197.2422 & 38.9185 & MDM & 3600 & 18.2$\pm$0.1 & 9.3$\pm$0.1 & IC~4182 & 321$\pm$4 & 274$\pm$36 & 47 \\
%MAGE~J1343+5813 & 205.7801 & 58.2278 & GBT & 2400 & 17.0$\pm0.1$ & 5.4$\pm$0.1 & NGC~5204 & 201$\pm$1 & 190$\pm$2 & 11  \\
\hline
\end{tabular}
}
\tablecomments{Column~1: ID-MAGE name Column~2: Right Ascension (J2000.0). Column~3: Declination (J2000.0). Column~4: Telescope observed with. Column~5: Exposure time in seconds for new observations. Column~6: Apparent $g-$band magnitude published in \citetalias{idmage}. Column~7: Effective radius in arcseconds published in \citetalias{idmage}.  Column~8: Host name. Column~9: Recessional velocity of the host in \kms. Column~10: Recessional velocity of the candidate in \kms\ from MDM observations. Column~11: Difference in the recessional velocities of the host and candidate in \kms. }
\label{table:velocity_confirmation}
\end{table*}

\subsection{MDM 2.4m}

We observed 73 satellite candidates with the MDM Observatory 2.4~m Hiltner telescope on Kitt Peak, during 2024 October 31--November 7 and 2025 March 24--30. We primarily targeted candidates with evidence of star formation based on GALEX NUV and/or FUV emission or blue optical colors (\textit{g$-$r}$\lesssim0.45$) to maximize the likelihood of detecting H$\alpha$ emission and, when possible, H$\beta$ and [\ion{O}{3}]~$\lambda\lambda 5007, 4958$ emission.  Because MDM long-slit observations require shallow imaging of the target for accurate slit placement, we prioritized candidates with brighter central surface brightness ($\mu_{0,g}\lesssim24.0$).  We used the Ohio State Multi-object Spectrometer (OSMOS; \citealt{OSMOS}) in single-slit mode with the ``blue'' disperser and the 3\arcsec\ ``inner'' slit.  The spectra have a wavelength coverage of 3980 to 6900 \r{A}, a dispersion of 0.7~\r{A} per 15 $\mu$m pixel, and a resolution (FWHM) of $\sim$7.6~\r{A}.  Targets were observed 3$\times$1200s in 2024 and 2$\times$1200s or 3$\times$1200s in 2025, depending on the quality of the first spectrum.

We used a Python script designed for OSMOS reduction available on github\footnote{https://github.com/jrthorstensen/thorosmos} to compute and subtract bias levels of the OSMOS detector's four amplifiers, apply the spectroscopic flat fields, and median combine the spectra. After standard spectroscopic reductions, we aperture-extracted the target and sky spectra in \emph{IRAF}. For the wavelength calibration we used a combination of Hg, Ne, and Xe comparison lamps taken at zenith during the day. To correct for the drift in the zero-point and dispersion (which can be significant at $\sim$4$-$5 \r{A}) of the science exposures, we used the 6300 and 5577 \r{A} night sky lines of the extracted sky spectra to improve the wavelength calibration.  The resulting calibration is accurate to $\leq 1$ pix or $\sim35$ \kms, sufficient to confidently separate satellite and background galaxies (Thorstensen et al. in prep.).  Due to the spectral lines having a boxy shape in the OSMOS data, we did not use a Gaussian fit to centroid the lines. Instead we identified the outer edges of the emission line and determined the line center from the mid-point of the broadened peak. OSMOS is optimized for the red, giving good sensitivity around 6500~\AA\ to 6900~\AA, but the sensitivity falls off quickly towards the blue end. Therefore, for almost all of the candidates, we measured velocities from H$\alpha$ emission and used [\ion{O}{3}] and H$\beta$ to confirm the measured velocity. For MAGE~J0738+1635, which has a velocity $>15000$~\kms, placing its H$\alpha$ line beyond the OSMOS wavelength range, we instead use the average velocity of its [\ion{O}{3}] and H$\beta$ lines.

Of the 73 galaxies observed with MDM, we detect emission lines and confidently measure redshifts for 36. Two candidates (MAGE~J1052+3646 and MAGE~J1308+3855) are confirmed as satellites based on their relative velocities ( {$\Delta v_{r} \lesssim 150$~\kms)} to their hosts, NGC~3432 and IC~4182, respectively.  Figures~\ref{3432_365} and \ref{4182_811} show the spectra with prominent emission lines marked.  Both candidates are in the DESI DR~1 catalog, and our MDM-derived velocities agree with the DESI measurements. The remaining 34 candidates with detected emission lines are background galaxies. Of these, eight have GBT \hi\ detections and twelve are in DESI DR1.

For the 37 candidates without emission line detections, about half were detected in continuum, but do not have sufficiently strong emission lines to measure velocities, and the rest were non-detections.  The non-detections were due to a combination of poor weather (multiple nights experienced at least partial cloud coverage), low-surface brightnesses, and failure to place the slit over an \ion{H}{2} or UV bright region where emission lines are strongest. We intend to re-attempt follow-up of these 37 candidates with a more sensitive telescope and instrument combination to either to uncover faint emission lines or to pursue deep photometry for SBF distances.

\subsection{SALT}

We observed five candidates with the Robert Stobie Spectrograph (RSS) on SALT in long-slit mode between 2025 January 4 and February 4 using a slit width of 1$\arcsec$.5. As with the MDM observations, we targeted candidates with evidence of star formation to maximize the likelihood of detecting H$\alpha$, H$\beta$, and [\ion{O}{3}] emission lines. Given the surface brightnesses of our targets ($\mu_{0,g}\gtrsim23.0$), a bright star was used to align the galaxy in the slit. The spectra cover 4763-6808~\AA\ with a spectral resolution of 0.645 \AA~per pixel. The wavelength range includes H$\alpha$ emission for candidates with recessional velocities $\lesssim11000$~\kms. We observed each candidate with 2 or 4 exposures per visibility track. The targets were dithered along the slit between exposures to improve sky subtraction.

SALT employs a reduction pipeline on delivery of the data that replaces bad pixels, corrects for auto gain, flat-fields the images, removes cosmic ray hits, fills the CCD gaps, applies wavelength calibrations, and rectifies all files on delivery \citep{pySALT}\footnote{\url{http://pysalt.salt.ac.za/}}. To complete the reduction, we adapted the remaining steps in Kevin Hainline's SALT long-slit data reduction pipeline \citep{Hainline13}\footnote{\url{https://mips.as.arizona.edu/~khainline/salt_redux.html}} in \emph{IRAF}. We then background subtracted and extinction corrected the 2-D spectra before aperture extracting the 1-D spectra and applying the barycentric correction. 

Due to the changing aperture of SALT as it tracks objects across the sky, spectra were extracted from individual exposures before being combined. For each spectrum, we normalized it to a continuum fit, aligned to the mean value at the red chip gap, and median-combined the 1-D spectra in order to obtain a final spectrum for each target. No absolute flux calibration was attempted. We then masked any remaining sky-line residuals and measured recessional velocities from prominent emission lines. 

For four of the five candidates, we detect H$\alpha$, H$\beta$, and the [\ion{O}{3}]~$\lambda\lambda 5007, 4958$ emission with high signal-to-noise. For these targets, we report the average velocity measured from the four lines, with uncertainties given by the standard deviation of the individual line velocities. One candidate, MAGE~J0611-3341, has no detected emission lines. Three candidates (MAGE~J0330-5120, MAGE~J0505-2452, and MAGE~J0608-3413) are clearly background and are removed from the ID-MAGE sample. MAGE~J0609-3225 has a recessional velocity of 713$\pm$11~\kms, within $\simeq$75~\kms\ of its assumed host ESO364-G29's velocity (787$\pm3$~\kms), making it a velocity-confirmed satellite. The SALT spectra for MAGE~J0609-3225 is plotted in Figure~\ref{0609-3225} with its prominent emission lines labeled.

\subsection{DESI Archival data}

As part of our follow-up, we cross-matched the ID-MAGE candidate list with the public DESI DR1  data (\citealt{desi}). DESI DR1 contains the first 13 months of public data and the release of the observed  {optical} spectra and spectral fit through the DESI Legacy Survey viewer\footnote{https://www.legacysurvey.org/viewer/desi-dr1}. To identify DESI counterparts, we queried the DESI redshift catalog for all objects with spectra within 12\arcsec\ of each candidate and required a conservative fit-reliability cut of $\Delta \chi^2\geq40$. The $\Delta \chi^2\geq40$ represents the statistical significance of the best-fitting spectral class relative to the next best fit; larger values correspond to more reliable spectral fits and redshift measurements \citep{desi}. For all matches, we compared the DESI fiber position and the candidate location using the Legacy Survey viewer and inspected the DESI spectra to ensure a good spectral fit. 

In total, we identified 45 candidates with reliable redshifts in the DESI catalog. Three candidates (MAGE~J1052+3628, MAGE~J1052+3646, and MAGE~J1308+3855) are confirmed to be satellites and 42 are found to be background galaxies.  For the candidates with velocities from MDM (twelve candidates) and/or GBT (five candidates), the DESI velocities agree within their uncertainties. The only exception is MAGE~J1240+6225 discussed in Section~\ref{green_bank}.

\subsection{Sample Revision and Current Status}

\begin{figure*}[!tb]
    \centering
    \includegraphics[width=\linewidth]{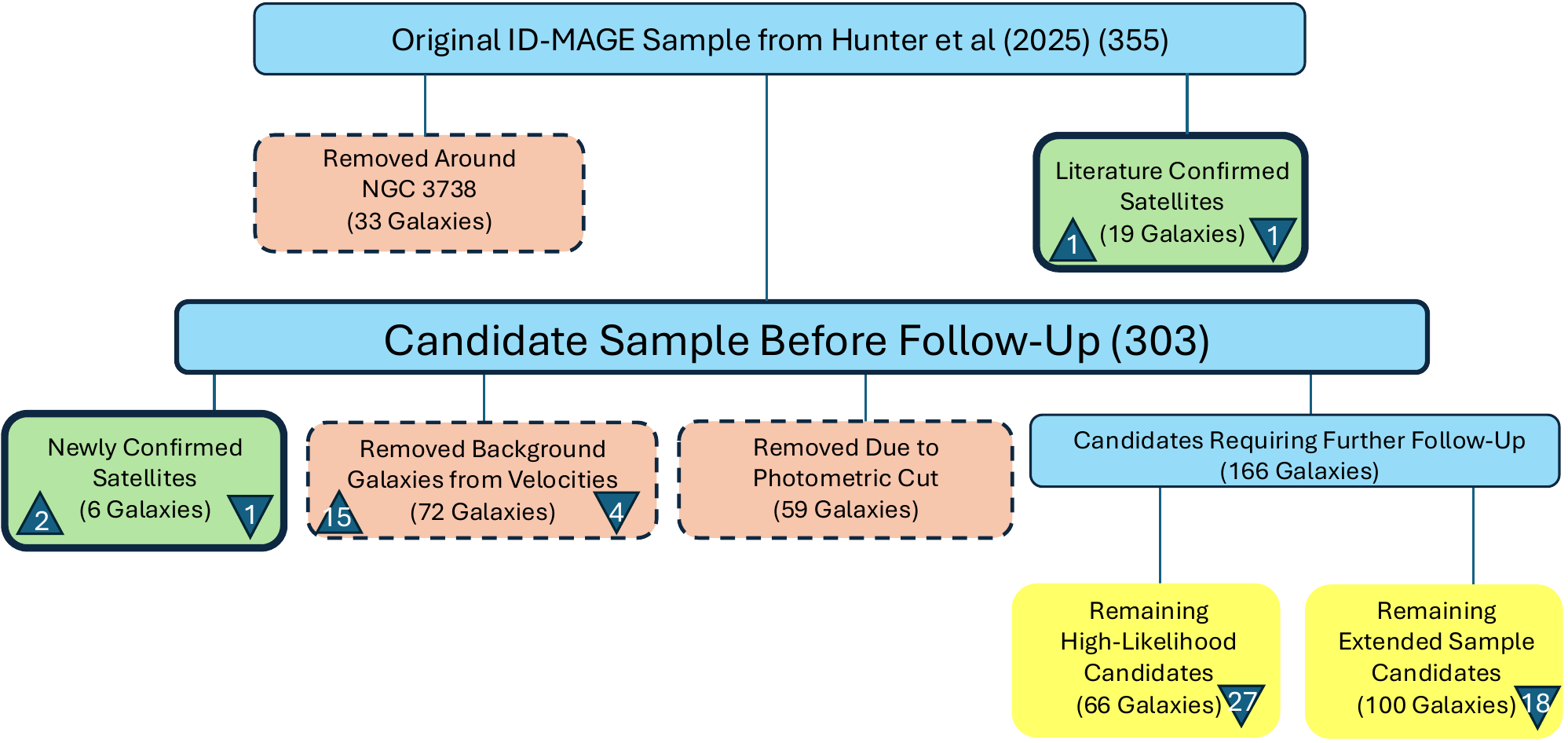}
    \caption{Decision tree summarizing the construction of the satellite and satellite candidate samples  {beginning with the sample identified in \citetalias{idmage}}. Blue boxes with solid borders indicate categories that are subsequently subdivided, orange boxes with dashed borders indicate candidates removed from the sample,  {yellow boxes with no borders are candidates requiring furthe follow-up observations,} and green boxes with bold borders indicate they  {are part of the Confirmed Satellite Sample}.  {The six Newly Confirmed Satellites and \background\ Removed Background Galaxies have velocity measurements from SALT, MDM, GBT, and/or DESI presented in this paper. Symbols within the boxes denote new GBT \hi\ data with} upward-pointing triangles indicate samples containing galaxies with new GBT \hi\ detections, while downward-pointing triangles indicate samples containing galaxies with new GBT \hi\ flux limits.  {The white number within the triangles denote the number of galaxies with this data available in Tables~\ref{table:satellite_HI}, \ref{table:Background_galaxies}, or \ref{table:GBT_limits}.} }
    \label{fig:decision}
\end{figure*}

In Tables~\ref{table:velocity_confirmation} and \ref{table:Background_galaxies} we present new velocities for \followedup\ satellite candidates identified in \citetalias{idmage}. { Based on the candidates' velocities, we confirm six candidates as satellites (lower green box in Figure~\ref{fig:decision}) and identify the remaining \background\ as background galaxies (first orange box in the fourth row of Figure~\ref{fig:decision}). All confirmed satellites were initially identified in the High Likelihood sample, and the background galaxies were initially from both samples (44 Extended, 28 High Likelihood).}

Among the \background\ rejected candidates, we identify five galaxies that are not satellites but may be blue, star-forming dwarf galaxies within $\simeq$15~Mpc (see Appendix Section~\ref{LV_dwarfs}). One, MAGE~J1250+5056, is part of the ELVES-Dwarf sample as a satellite of NGC~4707 (D$=$6.38$\pm$0.29~Mpc). It has an SBF distance of 7.8$^{+1.88}_{-1.31}$~Mpc \citep{ELVES-Dwarf}; however, NGC~4707 and MAGE~J1250+5056 have a  {$\Delta v_{r} \simeq225$~\kms} (Table~\ref{table:Background_galaxies}), well above our velocity threshold to be considered a satellite. Thus, we consider MAGE~J1250+5056 to be a background galaxy.

Similar to \cite{SAGAII}, we use our new spectroscopic catalog to improve our photometric selection criteria.  We find that, as expected, the Extended Sample has a higher background contamination fraction compared to the High-Likelihood Sample. In \citetalias{idmage}, only one previously confirmed satellite (MAGE~J1752+7008) was assigned to the Extended Sample, and none of the six newly velocity-confirmed satellites belong to this sample. Two of the five potentially Local Volume dwarf galaxies (see Appendix) are from the Extended Sample. Of the 44 Extended Sample candidates we have followed-up, we identify no galaxies that may be within $\simeq15$~Mpc and have effective radius r$_e \leq 5 \arcsec$.

Our spectroscopic campaign so far covers galaxies with colors of \textit{g-i$\leq$0.6} with only 2 candidates followed-up with \textit{g-r$>$0.6}.  This color preference is because we primarily targeted candidates with evidence of star formation for emission line and \hi\ spectroscopic follow-up. We therefore consider Extended Sample candidates with r$_e \leq 5 \arcsec$ and a color of \textit{g-r}$<$0.6 extremely unlikely to be satellites and no longer include them in our follow-up program. This removes \photcut galaxies from the ID-MAGE candidate sample. We continue to follow-up the remaining \fullcands\ Extended Sample candidates (i.e. those with $r_e>5\arcsec$ or \textit{g-r}$>$0.6). Based on the results of our follow-up observations, we consider the remaining Extended Sample galaxies to be galaxies of interest, although they are unlikely to be satellites of the targeted hosts. As such, we continue to prioritize follow-up of the High-Likelihood Sample candidates.

The updates to the ID-MAGE satellite sample presented in this paper are summarized in Figure~\ref{fig:decision}. This figure visualizes when candidates were removed from the sample, the current state of our follow-up campaign, and the data available for each subsample.  The current sample comprises 25 confirmed satellites (the six newly confirmed here and 19 previously confirmed satellites), and 166 unconfirmed satellite candidates (\highcands\ High-Likelihood Sample candidates, and \fullcands\ Extended Sample candidates). Table~\ref{tab:confirm} lists the 25 confirmed satellites along with the confirmation method, their velocities and/or distances, and the citation for the satellites' most reliable velocity or distance.

With the data presented here, we have completed follow-up observations of all High-Likelihood candidates for three hosts (two SMC-mass and one LMC-mass; see Section~\ref{completehosts}). Follow-up observations of one High-Likelihood candidate each will complete the satellite census for an additional 11 hosts. We are continuing follow-up observations of the \totcands\ unconfirmed candidates in ID-MAGE. 

\section{Discussion} \label{discussion}

With our significant progress in following-up the ID-MAGE candidates, we refine our qualitative estimates of the satellite population per host, and begin to characterize individual satellites and satellite systems.  In this section, we update our satellite population per host estimates, discuss the hosts we have completed follow-up for, and analyze the star formation properties and \hi\ gas content of our Confirmed Satellite Sample.

\begin{table*}
\caption{Confirmed Satellite Compilation}
\hspace*{-2.5cm}
\resizebox{1.13\linewidth}{!}{
\begin{tabular}{l r r r r l r l r }
\hline
\hline
Name & Alt. Name & RA & Dec & Host & Method & Sat Dist & Sat Vel. & Citation \\
 & & J2000 & J2000 & & & Mpc & \kms\ &  \\
(1) & (2) & (3) & (4) & (5) & (6) & (7) & (8) & (9) \\
\hline
MAGE J0006-2456 & ESO472-G15 & 1.7019 & -24.9446 & NGC~0024 & Vel &  & 570$\pm$2 & GBT: Table~\ref{table:satellite_HI}  \\
MAGE J0144+2717 & AGC~111945 & 26.1776 & 27.2889 & IC~1727/NGC~0672 & TRGB & 6.87$\pm$0.45 &  & T23 \\
MAGE J0155+2757 & AGC~111977 & 28.8355 & 27.9541 & NGC~0784 & TRGB & 5.50$\pm$0.11 &  & T23 \\
MAGE J0200+2849 &  AGC~111164 & 30.0428 & 28.8305 & NGC~0784 & TRGB & 5.11$\pm$0.07 &  & M14 \\
MAGE J0609-3225 & & 92.2913 & -32.4228 & ESO364-G29 & Vel. &  & 713$\pm$11 & SALT: Table~\ref{table:velocity_confirmation} \\
MAGE J0742+1633 & AGC 171379 & 115.6332 & 16.5613 & UGC~03974 & TRGB & 7.81$\pm$0.09 & 279$\pm$1 & T23, HKK \\
MAGE J1008+7038 & dw1008p7038b & 152.2005 & 70.6456 & UGC~05423 & SBF & 10.63$^{1.49}_{1.16}$ &  & ED \\
MAGE J1009+7032 & dw1009p7032 & 152.3944 & 70.5486 & UGC~05423 & SBF & 9.19 $^{0.76}_{0.66}$ &  & ED \\
MAGE J1052+3628 & LV J1052+3628 & 163.0238 & 36.4773 & NGC~3432 & Vel. &  & 481$\pm$10 & DESI: Table~\ref{table:velocity_confirmation} \\
MAGE J1052+3635 & UGC~05983 & 163.0723 & 36.5932 & NGC~3432 & Vel. &  & 699$\pm$33 & SDSS DR13  \\
MAGE J1052+3646 & & 163.1554 & 36.7695 & NGC~3432 & Vel. &  & 570$\pm$10 & DESI: Table~\ref{table:velocity_confirmation} \\
MAGE J1056+3608 & AGC~205685 & 164.1689 & 36.1392 & NGC~3432 & Vel &  & 572$\pm$3 & Yu22 \\
MAGE J1228+2235 & UGC~07584 & 187.0131 & 22.5881 & NGC~4455 & Vel. &  & 602$\pm$3 & Yu22 \\
MAGE J1228+2217 & AGC~223254 & 187.0212 & 22.2916 & NGC~4455 & TRGB & 6.39$\pm0.34$ &  & K20 \\
MAGE J1228+4358 & NGC~4449b & 187.1878 & 43.9711 & NGC~4449 & Vel. &  & $\simeq$230 & A23 \\
MAGE J1230+2312b & AGC~229379 & 187.6425 & 23.2061 & NGC~4455 & TRGB & 6.78$\pm0.37$ & 616$\pm$11 & K20, ALFALFA \\
MAGE J1245+6158 & SMDGJ1245495+615810 & 191.4558 & 61.9695 & NGC~4605 & SBF & 4.72$^{0.56}_{0.61}$ &  & ED \\
MAGE J1246+5136 & UGC~07950 & 191.7337 & 51.6128 & NGC~4707 & Vel. &  & 497$\pm$2 & SDSS DR13\\
MAGE J1256+3439 & UGCA~309 & 194.0749 & 34.6563 & NGC~4861 & Vel. &  & 730$\pm$3 &Yu22 \\
MAGE J1259+3528 & AGC~223250 & 194.7545 & 35.4806 & NGC~4861 & Vel. &  & 700$\pm$3 & Yu22 \\
MAGE J1259+3456 & LEDA~101479 & 194.8056 & 34.9363 & NGC~4861 & Vel. & & 930$\pm$1 & SDSS DR13  \\
MAGE J1305+3812 & & 196.4781 & 38.2112 & IC~4182 & Vel.  &  & 404$\pm$2 & GBT: Table~\ref{table:velocity_confirmation} \\
MAGE J1308+3855 & & 197.2422 & 38.9185 & IC~4182 & Vel. &  & 270$\pm$10 & DESI: Table~\ref{table:velocity_confirmation} \\
MAGE J1343+5813 & LEDA~2576103 & 205.7801 & 58.2278 & NGC~5204 & Vel. &  & 165$\pm$33 & SDSS DR13  \\
MAGE J1752+7008 &[KK98] 242 & 268.2012 & 70.1392 & NGC~6503 & TRGB & 6.46$\pm$0.32 & -66$\pm$10 & K22, P22 \\
\hline
\end{tabular}
}
\tablecomments{
Column~1: ID-MAGE identifier Column~2: Alternate names  Column~3: the Right Ascension (J2000.0). Column~4: the Declination (J2000.0). Column~5: Host Name Column~6: Method used to confirm satellite is associated with the host. Column~7: Velocity in \kms. Column~8: Distance in Mpc. Column~9: Confirmation data source (A23: \cite{4449FAST}, ALFALFA: \cite{ALFALFA}, ED: \cite{ELVES-Dwarf}, HHK: \cite{Huchtmeier03}, K20: \cite{KimYJ2020}, K22: \cite{Karachen22}, M14: \cite{McQuinn14}, P22: \cite{Pustilnik22}, SDSS DR13: \cite{SDSS13}, T23: \cite{Tully23}, Yu22: \cite{Yu22})}
\label{tab:confirm}
\end{table*}

\subsection{Updated Satellite Population per Host}\label{detect}

Following the methods in \citetalias{idmage}, we derive updated upper and lower estimates of the satellite population per host using the 25 confirmed satellites and the \highcands\ High-Likelihood Sample candidates. For this analysis, we divide the hosts listed in Table~\ref{table:hosts} into two mass bins: 9 LMC-mass hosts and 26 SMC-mass hosts. The upper estimate represents the maximum possible number of satellites in our sample and assumes that all High-Likelihood candidates will be confirmed as satellites.  We therefore include all the Confirmed Satellite Sample and High-Likelihood Sample candidates within a projected radius of 150~kpc for LMC-mass hosts and 110~kpc for SMC-mass hosts.  This radius cut extends slightly beyond the estimated virial radii for galaxies with comparable masses \citep{BMP2021, ELVES-Dwarf}. We find an upper bound of \LMCLFerup\ (\SMCLFerup) satellites per LMC (SMC)-mass host. The uncertainties are the standard deviation of the number of satellites per host, divided by $\sqrt{N_{host}}$.  The upper bounds are lower than those reported for the LMC-mass (\LMCLFold) and SMC-mass (\SMCLFold) hosts in \citetalias{idmage} because of the removal of confirmed background contaminants.

\begin{figure*}[!th]
    \centering
    \includegraphics[width=\linewidth]{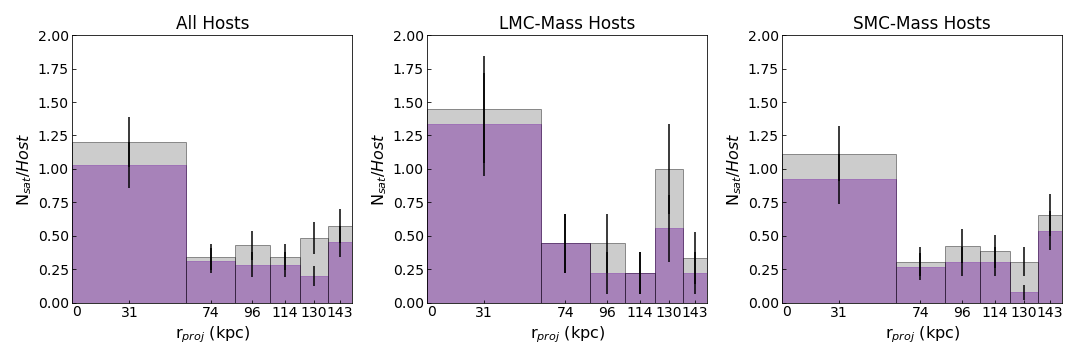}
    \caption{The central concentration of our confirmed satellites and High-Likelihood Sample candidates in the updated sample  {(purple)} compared to the corresponding sample from \citetalias{idmage} before the follow-up campaign began (gray): the number of galaxies per equal-area annular bin per host as a function of projected radius. Left panel: The sample for all hosts. Center panel: The sample for LMC-mass hosts. Right: The sample for SMC-mass hosts. Error bars represent the Poisson uncertainties in each bin. Projected distances are calculated using the distance of the assumed host. This equal-area binning highlights the concentration of galaxies within $\sim60$~kpc of their hosts. } 
    \label{fig:dist_host}
\end{figure*}

To account for the background contamination within the ID-MAGE sample, we use the spatial distribution of the satellites and candidates to obtain a more realistic estimate of the satellite population around our host galaxies. For this lower estimate of the satellite population, we spatially bin the satellites and candidates into annuli with equal projected area. The spatial distribution of the confirmed satellites and High-Likelihood candidates is shown in Figure~\ref{fig:dist_host} with the total satellite and candidate population (left), those around LMC-mass hosts (center), and those around SMC-mass hosts (right). The current sample is shown in purple, while the sample from \citetalias{idmage} is shown in gray for comparison. Each equal-area bin in Figure~\ref{fig:dist_host} should contain a roughly equal number of background galaxy contaminants because the number should depend primarily on the projected area. Figure~\ref{fig:dist_host} shows consistently fewer objects per radial bin in the updated sample than in \citetalias{idmage}, reflecting the removal of 36 confirmed background galaxies from the High-Likelihood Sample.

Following the methods in \citetalias{idmage}, we assume the number of satellites correlates with distance from the host while the background galaxy count remains roughly constant in all bins. We estimate an upper limit on the background contamination rate using the four outermost bins in Figure~\ref{fig:dist_host}. We find 0.31$\pm$0.09 contaminants per host per bin for both the LMC- and SMC-mass hosts. We adopt the same assumptions as \citetalias{idmage} that approximately 60\% of satellites reside within 60~kpc of their host. This assumption is based on the models in \cite{Dooley17a, Dooley17}, which use the dark matter-only \textit{Caterpillar} simulation suite \citep{Griffin16}, combined with different stellar-halo mass relationships (SHMR) \citep{Moster13, Brook14, GK16}. The inner annular bin, which has an outer radius of $\simeq$60~kpc, has 1.3$\pm$0.38 (0.9$\pm$0.18) galaxies per LMC-mass (SMC-mass) host. Subtracting the background contamination rate and multiplying by 1.67, to account for the 40\% satellites expected to be outside of 60~kpc, yields our lower estimate of \lowlimlmup\ (\lowlimsmup) satellites per LMC-mass (SMC-mass) host. This is in excellent agreement with our estimates from \citetalias{idmage} of \lowlimlm (\lowlimsm) satellites per LMC-mass (SMC-mass) host. The previous estimates were derived using the same methods before spectroscopic follow-up observations had commenced, demonstrating we properly accounted for the background contamination in our previous estimates. 

\subsection{Completed ID-MAGE Hosts}\label{completehosts}

\begin{figure}
    \centering
    \includegraphics[width=\linewidth]{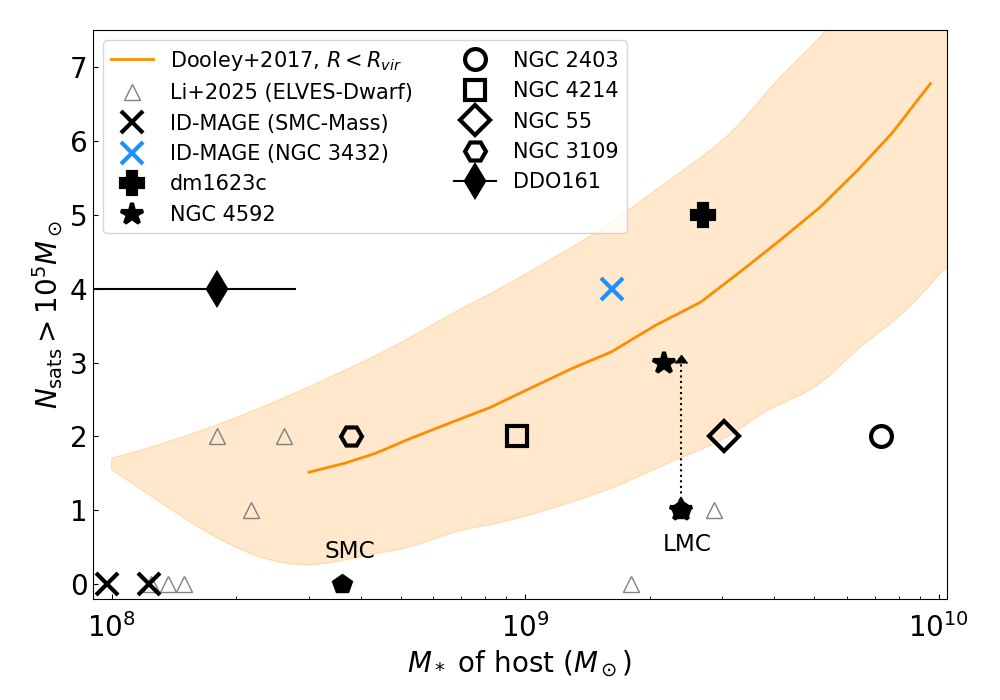}

    \caption{The total number of satellites based on the velocity-confirmed satellites around the ID-MAGE hosts NGC~3432, UGC~04426, and UGC~08201 (crosses) as a function of host stellar mass, compared with the predictions of \cite{Dooley17} (shaded orange). Confirmed satellite populations from previous systematic searches are included for comparison \citep{MADCASH, Carlin21, Carlin24, KimYJ2022, ADD3109, ELVES-Dwarf, Liddo161, Medoff25, stierwalt25}. For the LMC, a lower limit is used as its satellite population is uncertain due to the gravitational influence of the MW (e.g., \citealt{Pardy20, Patel20}. The stellar-mass uncertainty for DDO~161 is shown because it is significantly larger than those of the other hosts. Due to observational incompleteness, the satellite counts should be regarded as lower limits for most hosts. } 
    \label{fig:complete}
\end{figure}

We have completed follow-up observations of all High-Likelihood Sample candidates\footnote{The follow-up of Extended Sample candidates is ongoing.} for three ID-MAGE hosts (UGC~04426, UGC~08201, and NGC~3432) as the satellite candidates around these hosts are well-suited to our spectroscopic follow-up campaign. In figure~\ref{fig:complete}, we compare UGC~04426, UGC~08201, and NGC~3432's satellite populations with a sample of low-mass hosts from the literature that have volume-complete satellite searches.   %To compare their satellite populations, we compile a nearly complete literature sample of satellites bound to low-mass hosts that have been targeted by volume-complete satellite searches. 
Figure~\ref{fig:complete} shows the number of satellites as a function of host stellar mass for hosts within the mass range of our host sample, compared with the predictions of \citet{Dooley17a}, shown by the orange shaded region. Here, we require literature hosts to have at least $\gtrsim70\%$ of their virial volume surveyed with a well-defined completeness limit and all identified satellite candidates confirmed through velocity or distance measurements.  The satellite counts shown are lower limits for most hosts.  The majority of the satellite searches are not complete down to $M_\star=10^5 \msun$ (i.e., NGC~4592, ID-MAGE, dm1623c, ELVES-Dwarf, DDO~161; \citealt{KimYJ2022, idmage, stierwalt25, Liddo161, ELVES-Dwarf}). The resolved satellite searches around NGC~4214, NGC~2403 and NGC~3109 are complete to $M_\star\simeq10^5\msun$ but do not cover their full virial volumes \citep{MADCASH, Carlin21, Carlin24, ADD3109}.

Comparing the satellite populations of our completed hosts with previous observational studies reveals good overall agreement in the number of confirmed satellites around low-mass hosts \citep{Carlin21, Carlin24, KimYJ2022, ADD3109, ELVES-Dwarf, Liddo161, Medoff25}. Recent satellite searches have identified between zero and five satellites per low-mass host, with most systems hosting $\simeq0$--$2$ satellites and only a few hosting larger populations. The theoretical comparisons include predictions from \citet{Dooley17}, the semi-analytic \texttt{SatGen} framework \citep{SATGENI, SATGENII, ELVES-Dwarf}, and the TNG50 cosmological hydrodynamical simulation \citep{Pillepich18, pillepich19, Nelson19, Liddo161}. The \texttt{SatGen} predictions adopt different SHMRs \citep{Manwadkar22, Danieli23}, while the TNG50 predictions presented by \citet{Liddo161} use the empirical SHMR of \citet{Nadler20}. Overall, the satellite counts of our completed ID-MAGE hosts are consistent with both previous observations and theoretical expectations for hosts in this stellar-mass range.

\subsubsection{UGC~04426 and UGC~08201}

UGC~04426 and UGC~08201 are SMC-mass hosts with zero confirmed satellites and no likely candidates awaiting follow-up observations. There are three Extended Sample candidates in UGC~04426's FoV and one around UGC~08201, however; they are very unlikely to be satellites based on our confirmation success rate.  We therefore conclude that both hosts contain no satellites above our detection limit of $M_V$$\lesssim$$-9$. As seen in Figure~\ref{fig:complete}, previous studies have found three other SMC-mass hosts with no satellites within similar detection limits (\citealt{ELVES-Dwarf}). The absence of detected satellites around UGC~04426 and UGC~08201 is also inline with simulation predictions for $M_\star\simeq10^8~\msun$ hosts from semi-analytical models and simulations \citep{Dooley17, Santos-Santos22, ELVES-Dwarf, Liddo161}. Depending on SHMR, \texttt{SatGen} predicts $M_\star\simeq10^8~\msun$ hosts will have 0$^{+1}_{-0}$ satellites or 1$^{+2}_{-1}$ satellites with $M_\star>10^{5.6}~\msun$ \citep{ELVES-Dwarf}.  Using the results of TNG50 and the empirical SHMR from \cite{Nadler20}, \cite{Liddo161} predict hosts with $M_\star\simeq10^8~\msun$ will have 0$^{+1}_{-0}$ satellites with $M_\star\geq10^{5.4}~\msun$.

\subsubsection{NGC~3432}\label{3432_dis}

In \citetalias{idmage}, we identified eight candidates around the LMC-mass host NGC~3432, five in the High-Likelihood Sample and 3 in the Extended Sample.  Based on the velocity measurements presented here, four candidates are velocity-confirmed as satellites (MAGE J1052+3635 [aka UGC~05983], MAGE~J1052+3628, MAGE J1052+3646, and MAGE~J1056+3608), while three are background galaxies. The remaining galaxy to follow-up is from the Extended Sample and is very likely a background galaxy. The four velocity-confirmed satellites all appear to be star forming (Table~\ref{table:UV_SFR}) and three of the four are likely gas-rich, while MAGE J1052+3646 does not have a reliable \hi\ observation (Table~\ref{table:satellite_HI}: see Section~\ref{UVSFR} and Appendix~\ref{conf_plots}).

With four satellites, the number of satellites NGC~3432 hosts is in good agreement with predictions for a $1.5\times 10^9 \msun$ host \citep{Dooley17,Santos-Santos22,Liddo161, ELVES-Dwarf} (see the blue cross in Figure~\ref{fig:complete}). For hosts with $M_\star\simeq10^{9.2}~\msun$, \texttt{SatGen} models presented by \citet{ELVES-Dwarf} predict 2$^{+1}_{-2}$ satellites with $M_\star^{sat}>10^{5.6}~\msun$ when adopting the SHMR of \cite{Danieli23}, and 3$^{+6}_{-2}$ satellites when adopting the SHMR of \cite{Manwadkar22}. Similarly, \citet{Liddo161}, using TNG50 together with the SHMR of \citet{Nadler20}, predict 2$^{+2}_{-1}$ satellites with $M_\star^{sat}\geq10^{5.4}~\msun$.

\begin{figure}
    \centering
    \includegraphics[width=\linewidth]{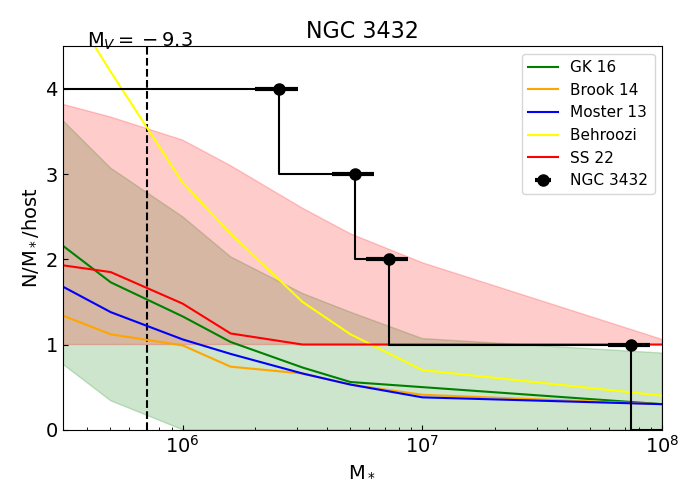}
    \caption{The satellite mass function for the LMC-mass host NGC~3432 in comparison to the \cite{Dooley17} (shaped green) and \cite{Santos-Santos22} (shaded red) model predictions. The vertical dashed line is the ID-MAGE completeness limit for NGC~3432.  The observed mass function for NGC~3432 is a $3\sigma$ outlier above the \cite{Dooley17} predictions and $1.5\sigma$ above \cite{Santos-Santos22}'s predictions.} 
    \label{fig:3432}
\end{figure}

Although the total number of satellites around NGC~3432 agrees with these predictions, their stellar-mass distribution is unusual. Figure~\ref{fig:3432} compares the satellite stellar mass function of NGC~3432 with the predictions from \cite{Dooley17} (green) and \cite{Santos-Santos22} (red).  The satellites have estimated stellar masses between $\log(M_*/M_\odot)\simeq6.4$ and 7.9, derived from their \textit{g-} and \textit{r-}band photometry using the relation from \cite{dlReyes24}.  NGC~3432's most massive satellite, MAGE~J1052+3635, has a stellar mass of $10^{7.9} M_\odot$, and is the second most massive satellite in the ID-MAGE sample.  

In Figure~\ref{fig:3432}, the mass function of NGC~3432's satellites is 3$\sigma$ above \cite{Dooley17}'s predictions and 1.5$\sigma$ above predictions from the ``cutoff'' model of \citet{Santos-Santos22}. \cite{Santos-Santos22} analyzed two models of the faint-end of the SHMR from recent high-resolution cosmological hydrodynamical simulations. The ``cut-off'' model is motivated by the APOSTLE simulations \citep{Fattahi16, Fattahi18} and disfavors galaxies forming in halos below the `hydrogen-cooling' limit (e.g., \citealt{Benitez-Llambay20}). In this model, the SHMR steepens at lower host masses, and hosts with M$_*\lesssim 3\times 10^8 \msun$ are expected to host no luminous satellites. 

 {Velocity confirmation may yield false positives, if one of NGC~3432's satellites} is a false positive, the system would shift from a 3$\sigma$ outlier to a 2$\sigma$ outlier for the \citet{Dooley17} predictions and into good agreement with the \citet{Santos-Santos22} predictions.  They require TRGB distance for more secure confirmation. At the NGC~3432's distance (8.93~Mpc), our satellite search is complete down to $M_V\simeq-9.3$ compared to our median sensitivity of $M_V\simeq-9$. As such, NGC~3432 may have additional satellites with $M_\star>10^5 \msun$, which would increase its satellite mass function in the lower mass range in Figure~\ref{fig:3432}. 

With four satellites, NGC~3432 has a richer satellite system than many hosts of similar stellar mass \citep{LBTsong, lbtsong2, Carlin24, ELVES-Dwarf}. As seen in Figure~\ref{fig:complete}, there are two other low-mass hosts with similarly rich satellite systems in the literature: dm1623c ($M_\star = 10^{9.43} \msun$), and DDO~161 ($M_\star = 10^{8.3} \msun$) \citep{ELVES-Dwarf, Liddo161, stierwalt25}. 
The two literature hosts have ``top-heavy" satellite mass functions compared to predictions \citep{Liddo161,stierwalt25}.  DDO~161's most massive satellite, UGCA~319, has a mass of $10^{7.50}~\msun$ and its other three satellites are $M_\star \simeq 10^{6.35}~\msun$. The satellites of dm1623c are more massive, with three having $M_\star>10^9~\msun$ ---roughly half the stellar mass of dm1623c. \cite{stierwalt25} suggests that dm1623c is part of a galaxy group (dm1623+15) comprised of multiple low-mass galaxies coming together to form a single system because of dm1623c's satellite mass function and the kinematics of the group. 

\cite{Liddo161} suggested that the satellite system of DDO~161 may represent a “too many satellites” problem, as it is a $3\sigma$ outlier relative to predictions from both TNG50 and their \texttt{SatGen} models. From TNG50, \cite{Liddo161} selected 857 isolated, central galaxies with stellar masses similar to DDO~161 and found a probability of less than 0.1\% that such a galaxy would host four satellites with $M_\star>10^{6.2}~\msun$.  However, as detailed in \cite{Liddo161}, DDO~161's stellar mass estimates range from $10^{7.91}~\msun$ \citep{deBlok2024} to $10^{8.5}~\msun$ (\citealt{Karachen13}), a substantially larger uncertainty than the $\simeq20\%$ uncertainties typical of the other hosts considered here. The resulting stellar mass ratio between DDO~161 and UGCA~319 ranges from 2.5 to 10, making it unclear if they are a host-satellite system or closer to equal-mass companions or a dwarf galaxy group like dm1623+15. To fully characterize unexpectedly rich satellite systems, such as NGC~3432 and DDO~161, and assess how they deviate from simulation predictions, further observational studies are required.

\subsection{Star Formation Rates and Gas Content of the Confirmed Satellite Sample}\label{UVSFR}

\begin{table*}
\caption{UV Star Formation Rates of the Confirmed Satellite Sample}
\hspace*{-2.5cm}
\resizebox{1.13\linewidth}{!}{
\begin{tabular}{l r r r r r r r r r}
\hline
\hline
Name &  RA & Dec & log(M$_\star$) &m$_{NUV}$ & log(SFR$_{NUV}$) & $\log(sSFR_{NUV}$) & m$_{FUV}$ & log(SFR$_{FUV}$) & $\log(sSFR_{FUV}$) \\
 & J2000 & J2000 & M$_\odot$ & mag & $\msun$\ yr$^{-1}$ & yr$^{-1}$ & mag & $\msun$\ yr$^{-1}$ & yr$^{-1}$  \\
(1) & (2) & (3) & (4) & (5) & (6) & (7) & (8) & (9) & (10) \\
\hline
MAGE~J1256+3439 & 194.0749 & 34.6563 & 8.0$\pm$0.2 & 17.0$\pm$0.2 & -2.25$\pm$0.09 & -10.3$\pm$0.2 & 17.5$\pm$0.2 & -2.63$\pm$0.09 & -10.6$\pm$0.2 \\
MAGE~J1052+3635 & 163.0723 & 36.5932 & 7.9$\pm$0.2 & 18.2$\pm$0.2 & -2.80$\pm$0.09 & -10.7$\pm$0.2 & 19.4$\pm$0.2 & -3.46$\pm$0.09 & -11.3$\pm$0.2 \\
MAGE~J1259+3456 & 194.8056 & 34.9363 & 7.7$\pm$0.2 & 19.2$\pm$0.1 & -3.13$\pm$0.04 & -10.9$\pm$0.2 & 19.4$\pm$0.1 & -3.39$\pm$0.04 & -11.1$\pm$0.2 \\
MAGE~J1246+5136 & 191.7337 & 51.6128 & 7.6$\pm$0.2 & 15.8$\pm$0.1 & -2.13$\pm$0.04 & -9.7$\pm$0.2 & 16.0$\pm$0.1 & -2.39$\pm$0.04 & -10.0$\pm$0.2 \\
MAGE~J0742+1633 & 115.6332 & 16.5613 & 7.6$\pm$0.2 & 17.6$\pm$0.2 & -2.66$\pm$0.09 & -10.2$\pm$0.2 & 17.9$\pm$0.1 & -2.96$\pm$0.04 & -10.5$\pm$0.2 \\
MAGE~J1228+2235 & 187.0131 & 22.5881 & 7.2$\pm$0.2 & 17.6$\pm$0.1 & -2.84$\pm$0.04 & -10.1$\pm$0.2 & 17.8$\pm$0.1 & -3.10$\pm$0.04 & -10.3$\pm$0.2 \\
MAGE~J0006-2456 & 1.7019 & -24.9446 & 7.2$\pm$0.2 & 18.7$\pm$0.1 & -3.02$\pm$0.04 & -10.4$\pm$0.2 & 19.2$\pm$0.1 & -3.58$\pm$0.04 & -10.8$\pm$0.2 \\
MAGE~J1228+4358 & 187.1878 & 43.9711 & 7.2$\pm$0.2 & $>$19.2 & $<$-3.88 & $<$-11.1 & $>$19.5 & $<$-4.16 & $<$-11.4 \\
MAGE~J1259+3528\tablenotemark{$\star$} & 194.7545 & 35.4806 & 7.1$\pm$0.2 & N/A & & & N/A &  &  \\
MAGE~J0155+2757 & 28.8355 & 27.9541 & 6.9$\pm$0.2 & 18.9$\pm$0.1 & -3.54$\pm$0.04 & -10.5$\pm$0.2 & 19.3$\pm$0.1 & -3.88$\pm$0.04 & -10.8$\pm$0.2 \\
MAGE~J1228+2217 & 187.0212 & 22.2916 & 6.9$\pm$0.2 & 18.2$\pm$0.1 & -3.08$\pm$0.04 & -10.0$\pm$0.2 & 18.5$\pm$0.2 & -3.38$\pm$0.09 & -10.3$\pm$0.2 \\
MAGE~J1052+3628 & 163.0238 & 36.4773 & 6.9$\pm$0.2 & 19.5$\pm$0.1 & -3.32$\pm$0.04 & -10.2$\pm$0.2 & 19.7$\pm$0.2 & -3.58$\pm$0.09 & -10.4$\pm$0.2 \\
MAGE~J0609-3225\tablenotemark{$\star$} & 92.2913 & -32.4228 & 6.8$\pm$0.2 & N/A & & &  N/A &  &  \\
MAGE J1245+6158 & 191.4558 & 61.9695 & 6.7$\pm$0.2 & 22.3$\pm$0.1 & -4.92$\pm$0.04 & -11.62$\pm$0.2 & $>22.1$ & $<-$5.1 & $<-$11.8 \\
MAGE~J1056+3608 & 164.1689 & 36.1392 & 6.7$\pm$0.2 & 18.5$\pm$0.2 & -2.92$\pm$0.09 & -9.6$\pm$0.2 & 18.6$\pm$0.2 & -3.14$\pm$0.09 & -9.9$\pm$0.2 \\
MAGE~J0144+2717 & 26.1776 & 27.2889 & 6.7$\pm$0.2 & 19.2$\pm$0.1 & -3.38$\pm$0.04 & -10.1$\pm$0.2 & 19.5$\pm$0.1 & -3.68$\pm$0.04 & -10.4$\pm$0.2 \\
MAGE~J0200+2849 & 30.0428 & 28.8305 & 6.6$\pm$0.2 & 19.6$\pm$0.2 & -3.82$\pm$0.09 & -10.4$\pm$0.2 & 19.9$\pm$0.2 & -4.12$\pm$0.09 & -10.7$\pm$0.2 \\
MAGE~J1009+7032 & 152.3944 & 70.5486 & 6.5$\pm$0.2 & 21.7$\pm$0.2 & -4.23$\pm$0.09 & -10.8$\pm$0.2 & 22.0$\pm$0.1 & -4.53$\pm$0.04 & -11.1$\pm$0.2 \\
MAGE~J1052+3646 & 163.1554 & 36.7695 & 6.4$\pm$0.2 & 21.1$\pm$0.1 & -3.96$\pm$0.04 & -10.4$\pm$0.2 & 21.2$\pm$0.1 & -4.18$\pm$0.04 & -10.6$\pm$0.2 \\
MAGE~J1343+5813 & 205.7801 & 58.2278 & 6.3$\pm$0.2 & 18.3$\pm$0.1 & -3.44$\pm$0.04 & -9.7$\pm$0.2 & 18.2$\pm$0.1 & -3.58$\pm$0.04 & -9.9$\pm$0.2 \\
MAGE~J1008+7038 & 152.2005 & 70.6456 & 6.2$\pm$0.2 & 20.7$\pm$0.2 & -3.83$\pm$0.09 & -10.0$\pm$0.2 & 20.6$\pm$0.3 & -3.97$\pm$0.14 & -10.2$\pm$0.2 \\
MAGE~J1752+7008 & 268.2012 & 70.1392 & 6.2$\pm$0.2 & 21.5$\pm$0.1 & -4.45$\pm$0.04 & -10.6$\pm$0.2 & 22.0$\pm$0.2 & -4.83$\pm$0.09 & -11.0$\pm$0.2 \\
MAGE~J1308+3855\tablenotemark{$\star$} & 197.2422 & 38.9185 & 6.2$\pm$0.2 & N/A & & &  N/A &  &  \\
MAGE~J1230+2312b & 187.6425 & 23.2061 & 6.1$\pm$0.2 & 20.9$\pm$0.1 & -4.16$\pm$0.04 & -10.3$\pm$0.2 & 21.2$\pm$0.2 & -4.46$\pm$0.09 & -10.6$\pm$0.2 \\
MAGE~J1305+3812 & 196.4781 & 38.2112 & 5.8$\pm$0.2 & 21.7$\pm$0.1 & -4.85$\pm$0.04 & -10.7$\pm$0.2 & 22.1$\pm$0.1 & -5.23$\pm$0.04 & -11.1$\pm$0.2 \\
\hline
\end{tabular}
}
\tablecomments{
\textit{$\star$} Outside of GALEX coverage
Column~1: ID-MAGE identifier Column~2: the Right Ascension (J2000.0). Column~3: the Declination (J2000.0). Column~4: stellar mass derived from Legacy Survey \textit{g} and \textit{r-}band data using the \cite{dlReyes24} calibration for low-mass galaxies. Column~5: Apparent GALEX NUV magnitude using optical effective radius. Column~6: SFR from NUV photometry. Column~7: Specific star formation rate (sSFR) from GALEX NUV photometry. Column~8: Apparent GALEX FUV magnitude using optical effective radius. Column~9: SFR from FUV photometry. Column~10: Specific star formation rate (sSFR) from GALEX FUV photometry.}
\label{table:UV_SFR}
\end{table*}

We use archival GALEX data to determine the star formation rates (SFRs) for the Confirmed Satellites Sample (Table~\ref{table:UV_SFR}).  Three (MAGE~J0609-3225, MAGE~J1259+3258, and MAGE~J1308+3855) fall outside of the GALEX footprint.  SFRs are derived from the NUV and FUV data within the GALEX All-Sky Survey and the Nearby Galaxies Survey \citep{galex2005}. We perform aperture photometry on the GALEX NUV and FUV tiles using the same process as \citet{jones2022}. We use circularized apertures defined by twice the galaxy’s half-light radius matching the apertures to those found from the optical photometry. Bright foreground sources and image artifacts are masked, and background noise uncertainties are estimated empirically by randomly placing 10,000 empty apertures across each GALEX tile, matched in area to the science aperture and restricted to regions free of extremely bright sources and edge effects. This estimation accounts for the statistical uncertainties in the UV fluxes, but not the systematic uncertainties, which we do not attempt to quantify. Fluxes are converted to AB magnitudes using the calibrations of \citet{morrissey2007}, and corrected for Milky Way (MW) extinction adopting $R_{\rm NUV} = 8.20$ and $R_{\rm FUV} = 8.24$ from \citet{wyder2007}. We do not correct for internal extinction as low-mass galaxies usually have minimal dust extinction \citep[e.g.,][]{mcquinn2015b}. De-reddened flux densities are then converted to luminosities using the TRGB distances to the satellite's host. SFRs are calculated following the procedure of \citet{iglesias2006}\footnote{This SFR conversion is calibrated for AB$_{NUV}~\leq~16$. However, it is routinely used for galaxies with AB$_{NUV}~\geq~16$ \citep{Karunakaran21, Jones24b, Fielder25}}, adopting a bolometric absolute magnitude of the Sun of 4.74, and are listed in Table~\ref{table:UV_SFR} with their GALEX photometry.

\begin{figure}
    \centering
    \includegraphics[width=\linewidth]{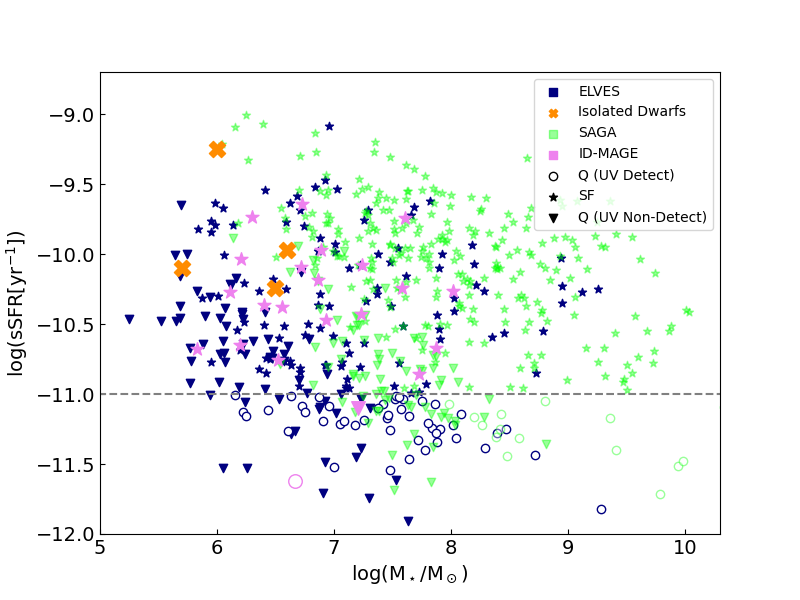}
    \caption{Specific star formation rate (sSFR$_{NUV}$) as a function of stellar mass for satellites in ELVES (navy), SAGA (green), and ID-MAGE (purple) samples and isolated dwarfs (orange). Star-forming satellites are defined as NUV-detected systems with $\log$(sSFR$_{NUV}$[yr$^{-1}$])$\geq$-11), are shown as stars and labeled SF in the legend.  Quenched satellites that are detected in the UV but have $\log$(sSFR$_{NUV}$[yr$^{-1}$])$<$-11 are shown as open circles. UV non-detections are shown as downward-pointing triangles representing upper limits.  Quenched systems are labeled Q in the legend.  {Star-forming isolated galaxies, are shown as x's and labeled Isolated Dwarfs in the legend.}  The dashed line is at $\log$(sSFR[yr$^{-1}$])$=$-11, the threshold between quenched and star-forming satellites.   The ELVES and SAGA data are compiled from \cite{ELVES, Karunakaran21, SAGAIII, SAGAIV}. }
    \label{fig:ssfr_mass}
\end{figure} 

We consider a satellite to be star forming if it is detected in the UV and has a NUV-specific star formation rate (sSFR = SFR/$M_{*}$) of $\log(sSFR[yr^{-1}])\geq-11$ \citep{Karunakaran23,SAGAIV}. Satellites with $\log(sSFR[yr^{-1}])<-11$ are considered quenched and not actively star forming. Of the confirmed ID-MAGE satellites, only two appear to quenched. MAGE~J1228+4358 (NGC~4449b) is a non-detection with GALEX, placing a limit of log(sSFR$_{NUV}$) $\leq-11.1$ yr$^{-1}$. Identified by \citet{Rich12} and \citet{MDelgado12}, the satellite is thought to be tidally disrupting as it closely interacts with its LMC-mass host, NGC~4449. The satellite has an estimated \hi-mass of $1.4\times10^7~\msun$ from \citeauthor{4449FAST}'s (\citeyear{4449FAST}) mapping of NGC~4449 with the Five-hundred-meter Aperture Spherical Telescope which is about 90\% of our stellar mass estimate in \citetalias{idmage} ($1.6\times10^7~\msun$). However, its \hi-mass is somewhat uncertain because it is within NGC~4449's disrupted and fragmented \hi\ disk.  The other quenched satellite is MAGE~J1245+6158, which is detected in the NUV with $\log(sSFR[yr^{-1}])=-11.62$  and is undetected in the FUV. No constraint on its \hi\ mass is currently available.  MAGE~J0609-3225, MAGE~J1259+3528, MAGE~J1308+3855 are outside of the GALEX footprint, but are unlikely to be quenched. All three have blue colors ($g-r\leq0.44\pm0.1$); MAGE~J0609-3225 and MAGE~J1308+3855 exhibit detectable H$\alpha$ emission, while MAGE~J1259+3528 has a high gas fraction ($M_{\rm HI}/M_\star=4.2$). 

Overall, the NUV SFRs and sSFRs of our satellites are comparable to the SFRs of galaxies within our mass range in different environments. The sSFRs for isolated galaxies reported in the literature span -9.3$\leq\log$(sSFR[yr$^{-1}$])$\leq$-10.3:
\setlist{nolistsep}
    \begin{itemize}[noitemsep]
    \item Leo~P has log(M$_\star/\msun)$$=$$5.7$$\pm$$0.1$) and $\log{(\rm{SFR}/\msun \rm{yr}^{-1})}$$=$$-4.4$) \citep{mcquinn2015b}
    \item Corvus~A has log(M$_\star/\msun)$$=$$6.0$$\pm0.1$), $\log{(\rm{SFR}/\msun \rm{yr}^{-1})}$$=$$-3.25\pm0.07$ \citep{Jones24b}
    \item LEDA~486718 has log(M$_\star/\msun)$$=$$6.5$$\pm$$0.1$), $\log{(\rm{SFR}/\msun \rm{yr}^{-1})}$$=$$-3.74\pm0.03$ \citep{Fielder25}
    \item Sculptor~26 has log(M$_\star/\msun)$$=$$6.6$$\pm$$0.1$), $\log{(\rm{SFR}/\msun \rm{yr}^{-1})}$$=$$-3.37\pm0.13$ \citep{Fielder25}
\end{itemize}
Figure~\ref{fig:ssfr_mass} compares the sSFRS of ID-MAGE satellites to sSFRS of satellites around MW-mass hosts from SAGA and ELVES \citep{Karunakaran21, ELVES, SAGAIV}  {and dwarf galaxies in isolated environments}. The star-forming satellites in ID-MAGE have a similar distribution of sSFRs to the ELVES and SAGA satellites in the same mass range. The ID-MAGE sample contains fewer quenched satellites and NUV non-detections, reflecting the fact that our follow-up observations to date have preferentially targeted star-forming candidates.

\begin{table*}
\caption{\hi\ Observations of Confirmed Satellites}
\hspace*{-2.5cm}
\resizebox{1.13\linewidth}{!}{
\begin{tabular}{l c r r r l r r r r}
\hline
\hline
Name & RA & Dec & Host & Host Vel. & Sat Vel. & $\Delta v_r$ & \hi\ Flux & \hi-mass & Cite \hi \\
 & J2000 & J2000 &  & \kms\ & \kms\ &\kms\ & Jy \kms\ & M$_\odot$ & \\
(1) & (2) & (3) & (4) & (5) & (6) & (7) & (8) & (9) & (10) \\
\hline
MAGE~J0006-2456 & 1.7019 & -24.9446 & NGC~0024 & 554$\pm$1 & 570$\pm$2 & 16 & 0.32$\pm$0.06 & 4.0$\pm0.7\times10^6$ & New GBT \\
MAGE~J0144+2717 & 26.1776 & 27.2889 & IC~1727/NGC~0672 & 345$\pm$1/429$\pm$1 & 423$\pm$1 & 78/5 & 3.12$\pm$0.05 & 3.6$\pm0.5 \times  10^7$ & ALFALFA \\
MAGE~J0155+2757 & 28.8355 & 27.9541 & NGC~0784 & 198$\pm$1 & 207$\pm$1 & 9 & 0.85$\pm$0.05 & 6.0$\pm 0.4\times10^6$ & ALFALFA  \\
MAGE~J0200+2849 & 30.0428 & 28.8305 & NGC~0784 & 198$\pm$1 & 163$\pm$14 & 35 & 0.65$\pm$0.04 & 4.0$\pm0.3\times10^6$ & ALFALFA \\
MAGE~J0742+1633 & 115.6332 & 16.5613 & UGC~03974 & 272$\pm$1 & 278$\pm$19 & 6 & 3.43$\pm$0.06 & 4.9$\pm0.1\times10^7$ & ALFALFA  \\
MAGE~J1009+7032 & 152.3944 & 70.5486 & UGC~05423 & 347$\pm$1 &  &  & $<$0.082 & $<$1.4$\times10^6$ & New GBT \\
%MAGE~J1052+3628\textsuperscript{\#} & 163.0238 & 36.4773 & NGC~3432 & 616$\pm$4 & 481$\pm$10 & 135 &  &  & New GBT \\ 
MAGE~J1056+3608 & 164.1689 & 36.1392 & NGC~3432 & 616$\pm$4 & 567$\pm$30 & 49 & 2.57$\pm$0.07 & 4.8$\pm0.9\times10^7$ & ALFALFA\\
MAGE~J1052+3635\textsuperscript{\textdagger} & 163.0723 & 36.5932 & NGC~3432 & 616$\pm4$ & 698$\pm1$ & 72 & 31.9$\pm1.1$\textsuperscript{\textdagger} & 5.9$\pm1.1\times10^8$ & OSvD23\\
%MAGE~J1052+3646 & 163.1554 & 36.7695 & NGC~3432 & 616$\pm$4 & & & $<$0.071 & $<$1.3$\times10^6$ & New GBT \\
MAGE~J1228+2235 & 187.0131 & 22.5881 & NGC~4455 & 644$\pm$1 & 602$\pm$12 & 42 & 3.28$\pm$0.05 & 3.2$\pm0.3\times10^7$ & ALFALFA \\
MAGE~J1228+2217 & 187.0212 & 22.2916 & NGC~4455 & 644$\pm$1 & 603$\pm$10 & 41 & 1.16$\pm$0.04 & 1.1$\pm0.1\times10^7$ & ALFALFA  \\
MAGE~J1228+4358 & 187.1878 & 43.9711 & NGC~4449 & 207$\pm$4 & 230 & 23 & & 1.4$\times10^7$ & AZX23  \\
MAGE~J1230+2312b & 187.6425 & 23.2061 & NGC~4455 & 644$\pm$1 & 624$\pm$11 & 20 & 0.38$\pm$0.04 & 4.1$\pm0.6\times10^6$ & ALFALFA  \\
MAGE~J1246+5136 & 191.7337 & 51.6128 & NGC~4707 & 468$\pm$1 & 508$\pm$5 & 40 & 7.3$\pm$0.6 & 7.0$\pm0.9\times10^7$ & HR98 \\
MAGE~J1256+3439 & 194.0749 & 34.6563 & NGC~4861 & 835$\pm$1 & 731$\pm$24 & 104 & 6.34$\pm$0.05 & 1.4$\pm0.1\times10^8$ & ALFALFA \\
MAGE~J1259+3528 & 194.7545 & 35.4806 & NGC~4861 & 835$\pm$1 & 699$\pm$14 & 136 & 2.36$\pm$0.06 & 5.2$\pm0.2\times10^7$ & ALFALFA \\
MAGE~J1305+3812 & 196.4781 & 38.2112 & IC~4182 & 321$\pm$4 & 404$\pm$2 & 83 & 0.29$\pm$0.04 & 1.2$\pm0.2\times10^6$ & New GBT \\
MAGE~J1308+3855 & 197.2422 & 38.9185 & IC~4182 & 321$\pm$4 & & & $<$0.168 & $<$7.1$\times10^5$ & New GBT \\
MAGE~J1343+5813 & 205.7801 & 58.2278 & NGC~5204 & 2-1$\pm$1 & 190$\pm2$ & 11 & 0.82$\pm$0.16 & 5.0$\pm1.0\times10^6$ & New GBT \\
\hline
\end{tabular}
}
\tablecomments{
\textdagger MAGE~J1052+3635 (UGC~05983) is in close proximity to NGC~3432, as such the \hi-flux is more uncertain than reported in the literature.  
Column~1: ID-MAGE identifier. Column~2: the Right Ascension (J2000.0). Column~3: the Declination (J2000.0). Column~4: Host name. Column~5: Recessional velocity of the host in \kms. Column~6: Recessional velocity of the candidate in \kms\ measured from \hi\ spectra. Column~7: Difference in the recessional velocities of the host and candidate in \kms. Column~8: \hi\ flux or flux limit in Jy \kms. Column~9: \hi-mass or \hi-mass limit in M$_\odot$ assuming the host's distance. Column~10: Citation for satellite \hi-flux ALFALFA: \cite{ALFALFA}, AZX23: \cite{4449FAST}, HR98: \cite{Huchtmeier98}, New GBT: New \hi\ measurements published in this paper, OSvD23: \cite{Oneil23}}
\label{table:satellite_HI}
\end{table*}

To study the gas content of the ID-MAGE satellites, we supplement our new GBT observations with published \hi\ measurements of the 19 previously confirmed satellites. The largest source of archival \hi\ measurements is the Arecibo Legacy Fast ALFA (ALFALFA; \citealt{ALFALFA}) survey, which includes detections of 10 confirmed satellites. We reviewed all \hi\ detections to asses their reliability and to check for contamination from the host or nearby galaxies. The \hi\ flux of MAGE~J1052+3635 is particularly uncertain because of its close proximity to NGC~3432 and is more uncertain then is reported in the literature from Nan\c{c}ay telescope observations in \cite{Oneil23}. Existing \hi\ maps show that the gas distributions of NGC~3432 and MAGE~J1052+3635 overlap both spatially and in velocity \citep{Wilcots96}.  

Our new GBT observations and the archival radio data are presented in Table~\ref{table:satellite_HI}. Three undetected galaxies in our new GBT observations are confirmed satellites and we can place \hi\ flux limits for two of them. We cannot place \hi\ flux limits for MAGE~J1052+3646 because its spectrum is contaminated by emission from its host NGC~3432, which is just over 9\arcmin\ away, and overlaps the satellite in velocity space.

We calculate the \hi\ mass or mass limits for the confirmed satellites with \hi\ detections or flux limits using:
\begin{equation}
    M_{\rm HI} = 2.356\times10^5 D^2 S_{HI}\ M_\odot 
\end{equation}
following \citet{Haynes98}, where $D$ is distance in Mpc, assumed to be the host distance, and $S_{\rm HI}$ is the flux in Jy \kms.  

In total, \hisats\ confirmed satellites have \hi\ detections or limit measurements, six have neither, and one (MAGE~J1052+3628) has a clear detection but an \hi\ mass cannot be derived. The \hi-to-stellar mass ratios for the confirmed satellites range from less than one-fifth to ten, typical of galaxies with stellar masses between $10^6$ and $10^{8}$ M$_\odot$ in isolated and group environments \citep[e.g,][]{LTHINGS, VLAANGST,Beale2020, Karunakaran24, garling24}. High \hi-to-stellar mass ratios are common for isolated galaxies in this stellar mass range \citep{Huang12}. 

While most of the confirmed satellites appear to be star-forming, a significant fraction are \hi\ deficient. We find 13 satellites have gas fractions ($M_{\rm HI} > M_\star$) marking them as gas-rich.  Five are \hi\ deficient ($M_{\rm HI} < M_\star$), including both satellites with \hi\ non-detections for which the upper limits imply $M_{\rm HI} \leq 0.5 M_\star$.  Four of the five \hi-deficient satellites nevertheless show clear evidence of ongoing star formation. MAGE~J0006-2456, MAGE~J0155+2757, and MAGE~J1009+7032 have $\log(sSFR[yr^{-1}])\geq-11.0$ yr$^{-1}$, while MAGE~J1308+3855 has clear H$\alpha$ and [\ion{O}{3}] emission in its MDM spectrum (Figure~\ref{4182_811}).

Our current Confirmed Satellite Sample is biased towards star-forming galaxies at lower stellar masses (M$_\star\leq10^7~\msun$). The results presented here demonstrate that velocity confirmation using emission line spectroscopy is effective for star-forming and gas-rich satellites but is less effective for quenched galaxies. Similarly, \citet{SAGAIII} found that for satellites with M$_\star\leq 10^{7.5}~M_\odot$, spectroscopic follow-up is more complete for star-forming candidates than for quenched candidates.  To mitigate this bias, we are conducting a complementary imaging campaign to measure SBF distances for redder candidates that are more likely to be quenched. This will enable a more representative comparison between the ID-MAGE satellite population and dwarf galaxies in other environments.

\subsection{Properties of High Mass Satellites}\label{massive}

In this section, we compare the \hi\ and star formation properties of the nine most massive confirmed satellites ($M_\star$$\geq$$10^{7}~\msun$), together with satellites in this mass range around literature hosts with completed surveys, with those of satellites around MW-mass hosts and isolated galaxies with $10^7~M_\odot \leq M_\star\leq10^8~\msun$.

Our follow-up of the most massive satellites ($M_\star$$\geq$$10^{7}~\msun$) in our sample is nearly complete. Nine of eleven satellites have measured velocities or distances, and eight have \hi\ measurements. While our reliance on velocity confirmation via emission-line detection preferentially favors star-forming satellites overall, this bias cannot explain the pattern at this mass range: satellite candidates here are already overwhelmingly star-forming, and only one system, MAGE~J1228+4358, shows clear evidence of quenching. The two High-Likelihood Sample candidates with $M_\star>10^7~\msun$ still requiring follow-up, MAGE~J2138-5430 and MAGE~J0607-3352, show evidence of star formation. They both have blue colors (\textit{g-i$<$0.33}), and MAGE~J2138-5430 has \textit{GALEX} NUV and FUV emission (MAGE~J0607-3352 is outside the \textit{GALEX} footprint).

From the \hi-data available, six of nine ($\simeq$67\%) confirmed satellites in this mass range are \hi-rich and two of nine ($\simeq$22\%) are \hi-deficient. The two \hi-deficient satellites are MAGE J1228+4358 and MAGE J0006-2456. MAGE J1228+4358 is discussed in Sections~\ref{UVSFR}. MAGE J0006-2456 has an \hi-mass fraction of 0.2 and has clear evidence of ongoing star formation with $\log(sSFR_{NUV})=-10.4\pm0.2$. One the nine confirmed satellites, MAGE J1259+3456, lacks a robust \hi-mass as it is $\simeq$5\arcmin\ from NGC~4861, too close to accurately measure its \hi-flux with the GBT.

In the literature, the satellite systems of NGC~2403 ($M_\star\simeq10^{9.8}~\msun$, \citealt{Dooley17a}) and NGC~4214 ($M_\star\simeq10^{8.8-8.9}~\msun$, \citealt{BMP2021, ELVES-Dwarf}) are the only low-mass hosts with well-constrained \hi-masses and recent SFRs for their satellites with M$\star\geq10^7~\msun$ (both hosts are included in Fig~\ref{fig:complete}). NGC~4214's satellite, DDO~113, is \hi-deficient with an \hi-to-stellar mass ratio less than one-half \citep{VLAANGST, Garling2020}, and NGC~2403's satellite, DDO~44, is extremely \hi-deficient with an \hi-mass limit less than one-four-hundredth its stellar mass \citep{Oneil23}.  Including NGC~4214 and NGC~2403's satellites with our confirmed satellites in this mass range, $\simeq$55\% of satellites are \hi-rich and $\simeq$36\% are \hi-deficient.  This roughly 60/40 split is consistent with the results of \citet{Karaunkaran22, Jones24a}, who found that $M_\star\gtrsim 10^7~\msun$ marks the mass scale where satellites of MW analogs transition from gas-poor (M$_{\text \hi}$/M$_{\star}\leq1$) to gas-rich (M$_{\text \hi}$/M$_{\star}\geq1$).  However, while the gas fractions of the satellites resemble those of MW analog satellites, this similarity does not seem to extend to their quenched fractions. 

Based on NUV sSFRs, only MAGE~J1228+4358 appears quenched among the nine ID-MAGE confirmed satellites with $M_\star>10^7 \msun$, resulting in a quenched fraction of $\simeq$10\%.  Including the satellites of DDO~113 and DDO~44 the quenched fraction becomes $\simeq$25\%  as both are quenched \citep{Weisz11, carlin19, Garling2020}. A quenched fraction of 10\% to 25\% falls roughly between isolated galaxies and satellites of MW-mass hosts; for this mass range, observations and simulations find $\leq$5\% of isolated dwarfs are quenched, while $\sim$40$-$70\% of MW-analog satellites are quenched \citep[e.g.,][]{geha2012, karachen2013b, slater2014, Weisz15, Greene23, christensen2024, SAGAIV, Mercado25}. 

\citet{Jahn22} analyzed Feedback In Realistic Environments (FIRE-2; \citealt{Hopkins18}) cosmological zoom-in simulations of six isolated LMC-mass galaxies to explore the impact of environment on satellite evolution. They investigated the satellites' SFHs, quenching fraction, and gas stripping in different satellite mass bins. \cite{Jahn22} found 33\% of M$_\star\geq10^7~\msun$ satellites to be quenched, roughly in agreement with the upper range of our observed quenched fraction (25\%). However, their predictions in this mass range are uncertain due to a sample size of only six satellites.  Additionally, they find all their satellites with M$_\star\leq10^7~\msun$ to be quenched, which does not agree with our current results. As seen in Table~\ref{table:UV_SFR}, we find 13 confirmed, star-forming satellites in the mas range $5.8\pm0.2\leq\log(M_\star/M_\odot)\leq6.9\pm0.2$ around both LMC- and SMC-mass hosts. The existence of an observed population of star-forming satellites with M$_\star\leq10^7~\msun$ not seen in simulations demonstrates the need for further modeling of the environmental impact of low-mass galaxies on their satellites.

Considering the higher \hi-deficient fraction and quenched fraction compared to isolated galaxies in the same mass range, it is clear low-mass hosts impact the quenching of their satellites.  However, comparing to MW-mass hosts, it appears that low-mass hosts are less efficient at quenching their satellites.  Our results suggest that the environment around low-mass hosts is driving the evolution of their satellites, and unsurprisingly, the impact is weaker than that of high-mass hosts.  Additionally, some current simulations may be overestimating the quenching efficiency of LMC-mass hosts.  As we complete our follow-up campaign, we will be able to determine the quenched fraction and \hi\ deficiency of the less massive satellites.

\section{Conclusions} \label{conclusions}

Our survey, ID-MAGE, is enabling detailed analysis of how low-mass hosts impact the evolution of their satellite populations. This paper presents the velocities of satellite candidates identified in ID-MAGE (\citetalias{idmage}) from follow-up observations taken in 2024 and 2025, and an analysis of confirmed satellites. In total, we present 110 velocity measurements for 83 candidates from new MDM 2.4 m, SALT, and GBT observations, and archival DESI DR1 data. We find \background\  candidates are background galaxies, and six are velocity-confirmed satellites. Combined with 19 confirmed satellites from the literature, our current Confirmed Satellite Sample is 25 galaxies.  

With our refined candidate sample, we derive estimates for the satellite populations of LMC- and SMC-mass host galaxies using the same methods as in \citetalias{idmage}. We find \lowlimlm (\lowlimsm) satellites per LMC-mass (SMC-mass) host with M$_V\lesssim-$9, consistent with previous searches such as ELVES-Dwarf, MADCASH, and DELVE-DEEP, as well as predictions from \citet{Dooley17} and in excellent agreement with \citetalias{idmage}. 

We have completed follow-up observations of the candidates for three of our 35 hosts.  Two SMC-mass hosts (UGC~04426 and UGC~08201) have no identified satellites.  This is consistent with previous satellite searches that have found SMC-mass hosts without satellites \citep[e.g.,][]{ELVES-Dwarf}, as well as model predictions that some $M_\star\sim10^8~\msun$ galaxies will host no satellites within our detection limits ($M_V \leq -9$) \citep{Dooley17a, Santos-Santos22}.  

The third host with completed follow-up observations is NGC~3432, an LMC-mass host with four velocity-confirmed satellites. While the number of satellites hosted by NGC~3432 is in good agreement with predictions, the masses of its satellites are significantly above the predictions of \citet{Dooley17} and more closely match the LFs of \citet{Santos-Santos22}. %Three other low-mass galaxies in the literature (DDO~161, \citealt{Liddo161}; NGC~1156, \citealt{ELVES-Dwarf}; dm1623c, \citealt{stierwalt25}) 
Two other low-mass galaxies in the literature (DDO~161, \citealt{Liddo161}; dm1623c, \citealt{stierwalt25}) 
are known to have similarly rich satellite systems (Section~\ref{3432_dis}).  As we complete follow-up observations, we will better constrain the host-to-host scatter and occurrence rate of satellite systems with ``top-heavy'' mass functions such as NGC~3432.

In Sections~\ref{green_bank} and \ref{UVSFR}, we present \hi\ observations for \gbtobs\ ID-MAGE candidates, including \gbtdetect\ new detections, and \gbtnodetect\ non-detections which can be used to determine \hi\ flux limits. Based on the GBT observations and archival \hi\ observations, 13 confirmed satellites are gas-rich ($M_{\rm HI}/M_\star > 1$), and five are gas-poor ($M_{\rm HI}/M_\star < 1$). From GALEX NUV and FUV photometry, two (MAGE~J1228+4358, and MAGE~J1245+6158) have $\log(sSFR[yr^{-1}])\leq-11~yr^{-1}$, indicating they are quenched, and 20 satellites have evidence of ongoing star formation. As the follow-up presented so far has focused on candidates' velocities, there is a bias towards gas-rich and star-forming satellites for those with $M_\star<10^7 \msun$. 

Our follow-up of the most massive satellites ($M_\star$$\geq$$10^{7}~\msun$) in our sample is nearly complete. Nine of eleven satellites have measured velocities or distances, and eight have \hi\ measurements. While our reliance on velocity confirmation via emission-line detection preferentially favors star-forming satellites overall, this bias cannot explain the pattern at this mass range: satellites in ID-MAGE with $M_\star$$\geq$$10^{7}~\msun$ are overwhelmingly star-forming, and only one system, MAGE~J1228+4358, shows clear evidence of quenching. Combining our sample of confirmed satellites with those of NGC~2403 (DDO~44) and NGC~4214 (DDO~113), we find $\simeq$55\% are \hi-rich and $\simeq$36\% are \hi-deficient. This 60/40 gas-rich and gas-poor ratio is consistent with the results for $M_\star>10^7~\msun$ mass satellites of MW analogs.  \cite{Karaunkaran22} and \cite{Jones24a} found that satellites of MW-mass hosts in this mass range are transitioning between gas-poor and gas-rich populations. 

Considering the star-forming properties of satellites with M$_\star\geq10^7~\msun$, only MAGE~J1228+4358, DDO~44, and DDO~113 are confirmed to be quenched. We calculate a quenched fraction of 10$-$25\%, which falls between the few-percent quenched fraction observed for isolated dwarfs \citep{geha2012} and the $\geq$50\% quenched fraction for satellites of MW-mass hosts \citep{Greene23,SAGAIV}. The higher quenched fraction relative to isolated dwarfs, together with the nearly 50\% \hi\ deficiency rate, demonstrates that low-mass hosts do significantly impact the evolution of their satellites.  The lower quenched fraction compared to satellites of MW-mass hosts illustrates the importance of understanding how low-mass hosts specifically quench their satellites. 

The completed ID-MAGE survey will include a diverse sample of satellites around low-mass hosts, providing key insights into how this environment impacts a galaxy's properties. Our goal is to create a more complete picture of the formation and evolution of low-mass galaxies and their satellites within the $\Lambda$CDM context.

\begin{acknowledgments}

LCH thanks Prof. John Thorstensen for his assistance with MDM 2.4m OSMOS observations and reductions. We thank Profs. Brian Chaboyer, John Thorstensen, and Ryan Hickox, and the SALT staff for their assistance with obtaining and reducing the SALT data. \textbf{We also thank Nathaniel Thornell for his work updating the SALT analysis pipeline. }We thank the reviewer for  {their} thorough comments which have greatly improved the text of this paper.

BMP acknowledges support from NSF grant AST-2508745.
DJS and the Arizona team acknowledges support from NSF grant AST-2205863.  
DC acknowledges support from NSF grant AST-2508747.
DJS \& DZ acknowledge support from NSF grants AST-2508746. 
KS acknowledges funding from the Natural Sciences and Engineering Research Council of Canada (NSERC).
ADD acknowledges support from STFC grants ST/Y002857/1.

This material is based upon work supported by the National Radio Astronomy Observatory and Green Bank Observatory which are major facilities funded by the U.S. National Science Foundation operated by Associated Universities, Inc. 

This research has made use of the NASA/IPAC Extragalactic Database (NED), which is operated by the Jet Propulsion Laboratory, California Institute of Technology, under contract with NASA.

This research is based on observations made with the Galaxy Evolution Explorer, obtained from the MAST data archive at the Space Telescope Science Institute, which is operated by the Association of Universities for Research in Astronomy, Inc., under NASA contract NAS 5–26555.

Some of the observations reported in this paper were obtained with the Southern African Large Telescope (SALT) under program 2024-2-DDT-001 (PI: L. C. Hunter).

This work is based on observations obtained at the MDM Observatory, operated by Dartmouth College, Columbia University, Ohio State University, Ohio University, and the University of Michigan.

The Legacy Surveys consist of three individual and complementary projects: the Dark Energy Camera Legacy Survey (DECaLS; Proposal ID \#2014B-0404; PIs: David Schlegel and Arjun Dey), the Beijing-Arizona Sky Survey (BASS; NOAO Prop. ID \#2015A-0801; PIs: Zhou Xu and Xiaohui Fan), and the Mayall z-band Legacy Survey (MzLS; Prop. ID \#2016A-0453; PI: Arjun Dey). DECaLS, BASS and MzLS together include data obtained, respectively, at the Blanco telescope, Cerro Tololo Inter-American Observatory, NSF’s NOIRLab; the Bok telescope, Steward Observatory, University of Arizona; and the Mayall telescope, Kitt Peak National Observatory, NOIRLab. Pipeline processing and analyses of the data were supported by NOIRLab and the Lawrence Berkeley National Laboratory (LBNL). The Legacy Surveys project is honored to be permitted to conduct astronomical research on Iolkam Du’ag (Kitt Peak), a mountain with particular significance to the Tohono O’odham Nation.

NOIRLab is operated by the Association of Universities for Research in Astronomy (AURA) under a cooperative agreement with the National Science Foundation. LBNL is managed by the Regents of the University of California under contract to the U.S. Department of Energy.

This project used data obtained with the Dark Energy Camera (DECam), which was constructed by the Dark Energy Survey (DES) collaboration. Funding for the DES Projects has been provided by the U.S. Department of Energy, the U.S. National Science Foundation, the Ministry of Science and Education of Spain, the Science and Technology Facilities Council of the United Kingdom, the Higher Education Funding Council for England, the National Center for Supercomputing Applications at the University of Illinois at Urbana-Champaign, the Kavli Institute of Cosmological Physics at the University of Chicago, Center for Cosmology and Astro-Particle Physics at the Ohio State University, the Mitchell Institute for Fundamental Physics and Astronomy at Texas A\&M University, Financiadora de Estudos E Projetos, Fundacao Carlos Chagas Filho de Amparo, Financiadora de Estudos E Projetos, Fundacao Carlos Chagas Filho de Amparo a Pesquisa do Estado do Rio de Janeiro, Conselho Nacional de Desenvolvimento Cientifico E Tecnologico and the Ministerio da Ciencia, Tecnologia E Inovacao, the Deutsche Forschungsgemeinschaft and the Collaborating Institutions in the Dark Energy Survey. The Collaborating Institutions are Argonne National Laboratory, the University of California at Santa Cruz, the University of Cambridge, Centro de Investigaciones Energeticas, Medioambientales y Tecnologicas-Madrid, the University of Chicago, University College London, the DES-Brazil Consortium, the University of Edinburgh, the Eidgenossische Technische Hochschule (ETH) Zurich, Fermi National Accelerator Laboratory, the University of Illinois at Urbana-Champaign, the Institut de Ciencies de l’Espai (IEEC/CSIC), the Institut de Fisica d’Altes Energies, Lawrence Berkeley National Laboratory, the Ludwig Maximilians Universitat Munchen and the associated Excellence Cluster Universe, the University of Michigan, NSF’s NOIRLab, the University of Nottingham, the Ohio State University, the University of Pennsylvania, the University of Portsmouth, SLAC National Accelerator Laboratory, Stanford University, the University of Sussex, and Texas A\&M University.

BASS is a key project of the Telescope Access Program (TAP), which has been funded by the National Astronomical Observatories of China, the Chinese Academy of Sciences (the Strategic Priority Research Program “The Emergence of Cosmological Structures” Grant \# XDB09000000), and the Special Fund for Astronomy from the Ministry of Finance. The BASS is also supported by the External Cooperation Program of Chinese Academy of Sciences (Grant \# 114A11KYSB20160057), and Chinese National Natural Science Foundation (Grant \# 12120101003, \# 11433005).

The Legacy Survey team makes use of data products from the Near-Earth Object Wide-field Infrared Survey Explorer (NEOWISE), which is a project of the Jet Propulsion Laboratory/California Institute of Technology. NEOWISE is funded by the National Aeronautics and Space Administration.

The Legacy Surveys imaging of the DESI footprint is supported by the Director, Office of Science, Office of High Energy Physics of the U.S. Department of Energy under Contract No. DE-AC02-05CH1123, by the National Energy Research Scientific Computing Center, a DOE Office of Science User Facility under the same contract; and by the U.S. National Science Foundation, Division of Astronomical Sciences under Contract No. AST-0950945 to NOAO.

This research used data obtained with the Dark Energy Spectroscopic Instrument (DESI). DESI construction and operations is managed by the Lawrence Berkeley National Laboratory. This material is based upon work supported by the U.S. Department of Energy, Office of Science, Office of High-Energy Physics, under Contract No. DE–AC02–05CH11231, and by the National Energy Research Scientific Computing Center, a DOE Office of Science User Facility under the same contract. Additional support for DESI was provided by the U.S. National Science Foundation (NSF), Division of Astronomical Sciences under Contract No. AST-0950945 to the NSF’s National Optical-Infrared Astronomy Research Laboratory; the Science and Technology Facilities Council of the United Kingdom; the Gordon and Betty Moore Foundation; the Heising-Simons Foundation; the French Alternative Energies and Atomic Energy Commission (CEA); the National Council of Humanities, Science and Technology of Mexico (CONAHCYT); the Ministry of Science and Innovation of Spain (MICINN), and by the DESI Member Institutions: www.desi.lbl.gov/collaborating-institutions. The DESI collaboration is honored to be permitted to conduct scientific research on I’oligam Du’ag (Kitt Peak), a mountain with particular significance to the Tohono O’odham Nation. Any opinions, findings, and conclusions or recommendations expressed in this material are those of the author(s) and do not necessarily reflect the views of the U.S. National Science Foundation, the U.S. Department of Energy, or any of the listed funding agencies.

\textit{Facilities}: MDM, SALT, GBT, GALEX

\textit{Software:} Astropy \citep{astropy13,astropy18,astropy22}, GALFIT \citep{Peng_2010}, \emph{IRAF} \cite{IRAF86,IRAF93}, Matplotlib \cite{Hunter:2007}, NumPy \citep{harris2020array}, pandas \citep{mckinney-proc-scipy-2010}.

\end{acknowledgments}

\appendix

\section{Spectra of Velocity-Confirmed Satellites}\label{conf_plots}

\begin{figure*}
    \centering
    \includegraphics[width=.6\linewidth]{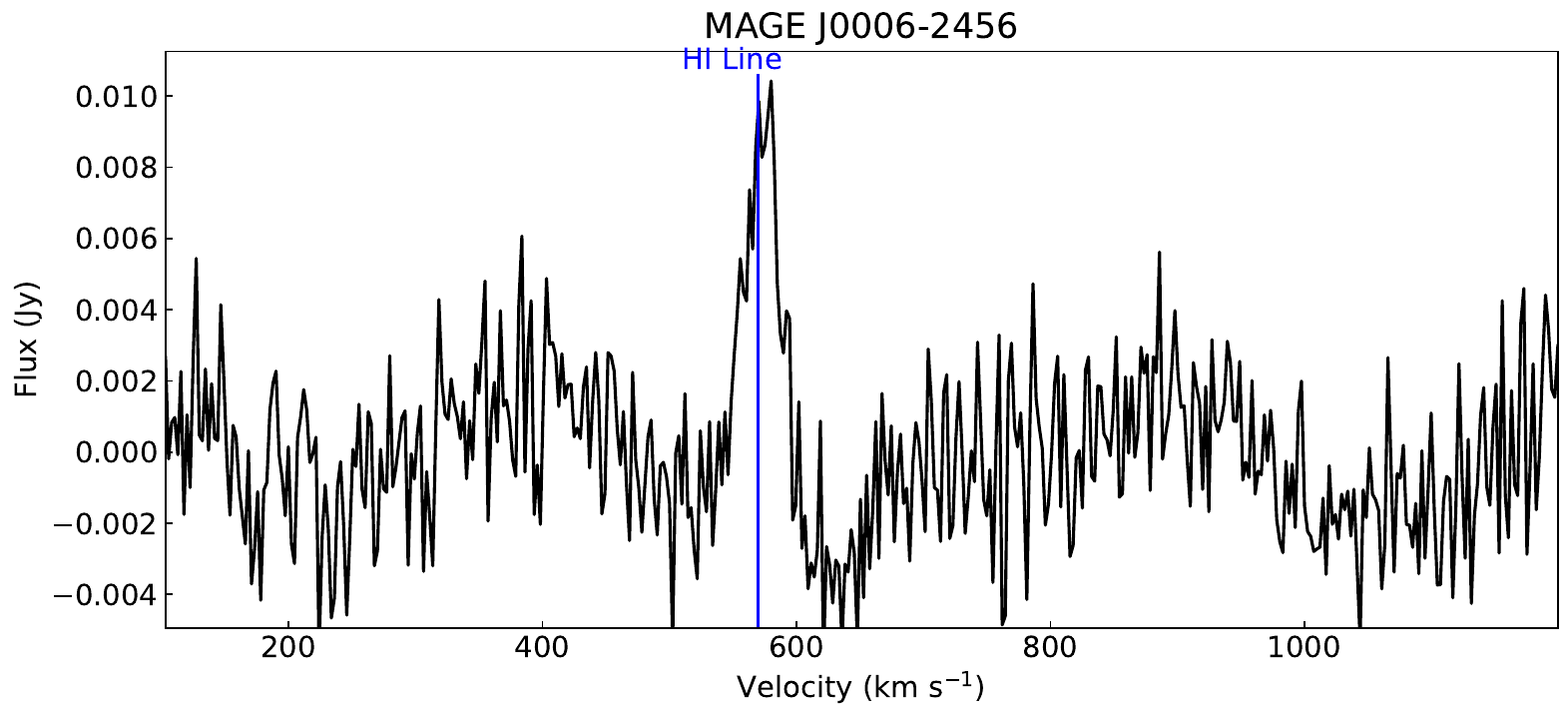}
    \caption{GBT \hi\ spectrum of MAGE~J0006-2456 taken as part of project ID GBT25A-155. The \hi\ emission line used to derive the velocity (570$\pm$2\kms) is marked.} 
    \label{0006-2456}
\end{figure*} 

\begin{figure*} 
    \centering
    \includegraphics[width=.55\linewidth]{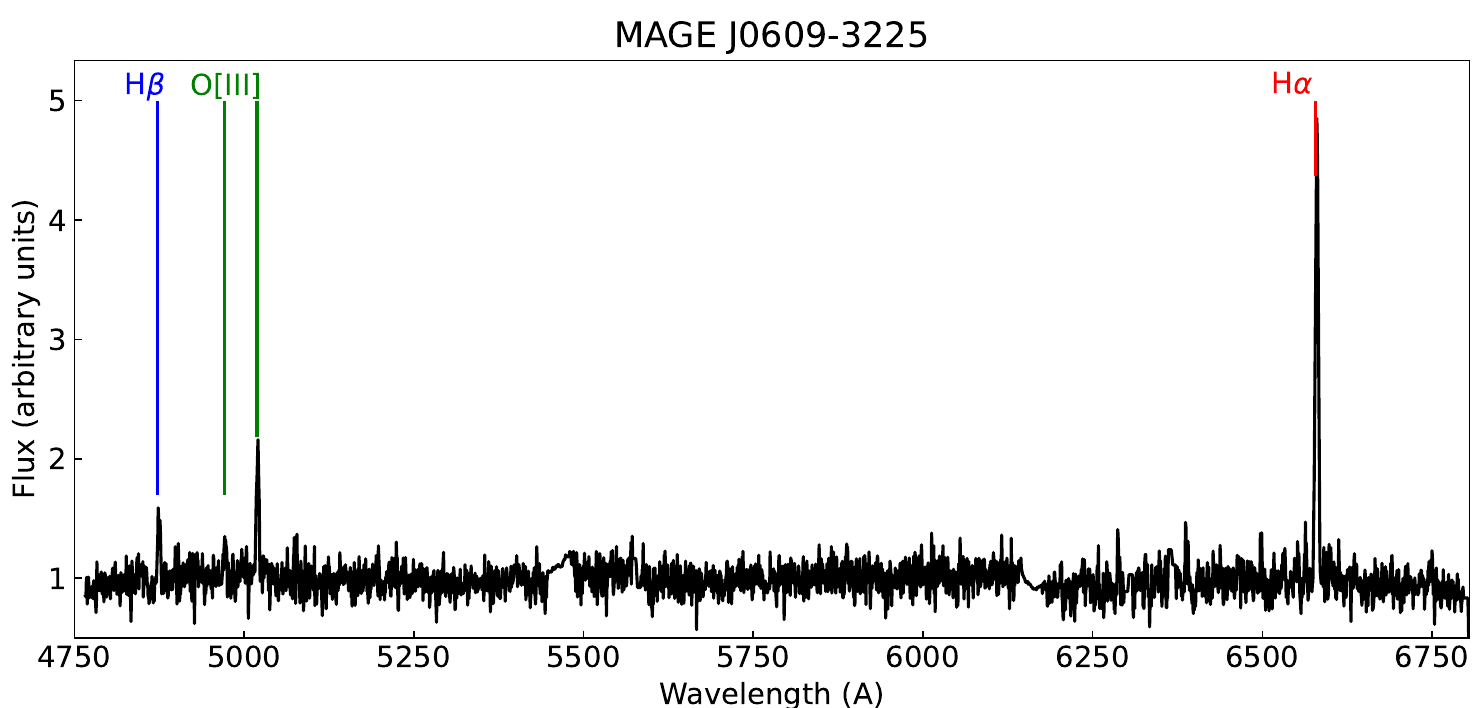}
    \caption{SALT spectrum of MAGE J0609-3225 taken January 2025. The H$\alpha$, H$\beta$, and [\ion{O}{3}] emission lines used to derive the velocity (713$\pm$40~\kms) are marked. The spectra are not flux calibrated. } 
    \label{0609-3225}
\end{figure*} 

\begin{figure*}
    \centering
    \includegraphics[width=.55\linewidth]{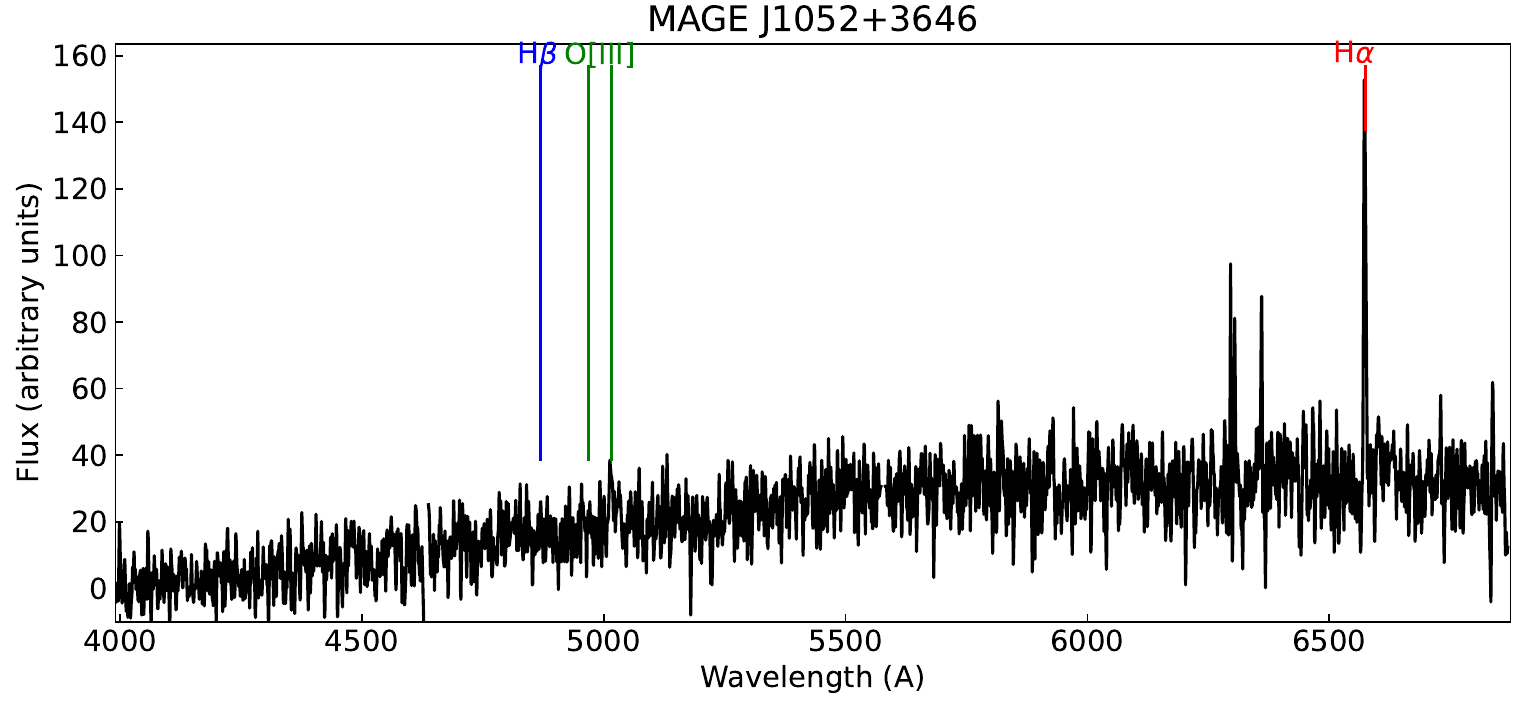}
    \caption{MDM 2.4~m spectrum of MAGE J1052+3646 taken April 2025. The H$\alpha$, H$\beta$, and [\ion{O}{3}] emission lines used to derive the velocity (523$\pm$40\kms) are marked. The spectra are not flux calibrated. } 
    \label{3432_365}
\end{figure*} 

\begin{figure*}
    \centering
    \includegraphics[width=.6\linewidth]{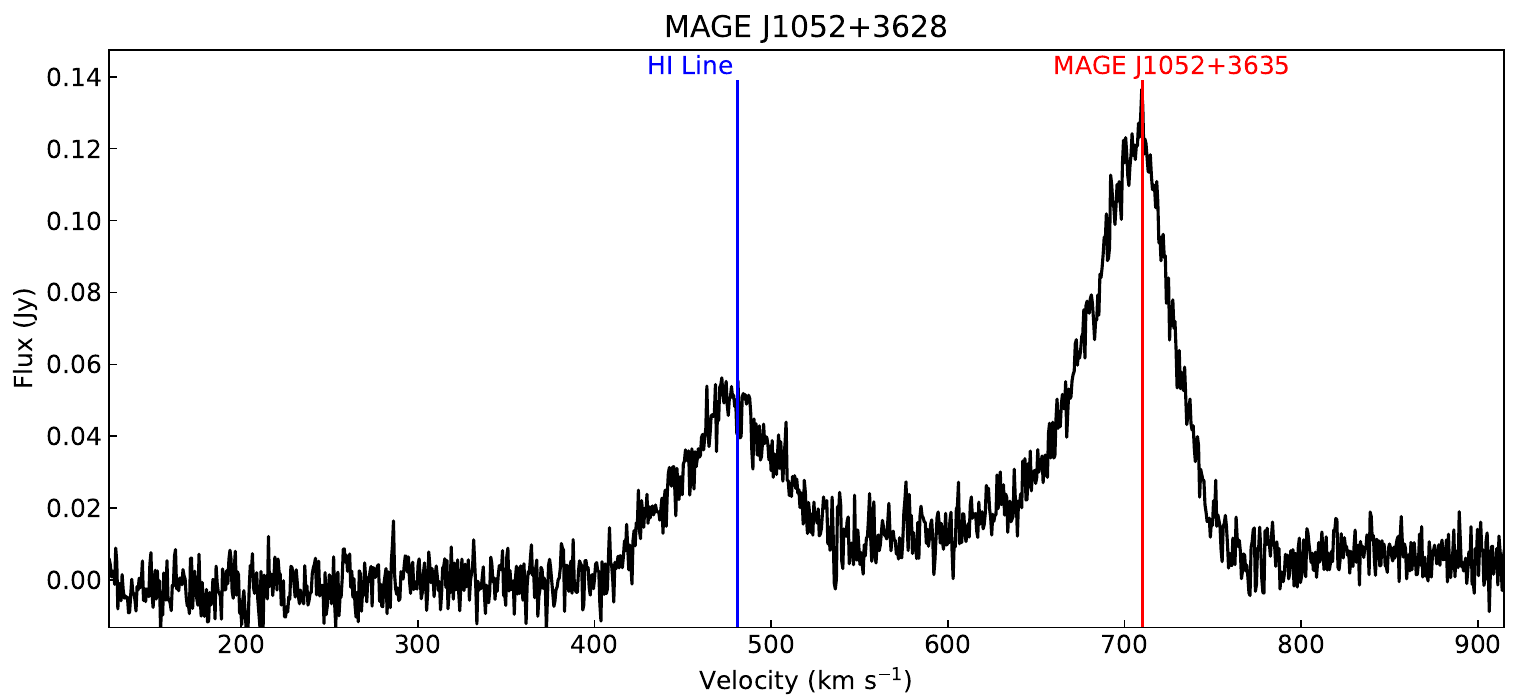}
    \caption{GBT \hi\ spectrum of MAGE J1052+3628 taken as part of project ID GBT25A-155. The \hi\ emission line used to derive the velocity (481$\pm$2\kms) is marked. Contamination from UGC~05983, another velocity-confirmed satellite of NGC~3432, can be seen at $\simeq$700\kms. Additionally, contamination from the host NGC~3432 is evident in the elevated baseline between 450~\kms\ and 700~\kms. } 
    \label{1052+3628}
\end{figure*} 

\begin{figure*}
    \centering
    \includegraphics[width=.6\linewidth]{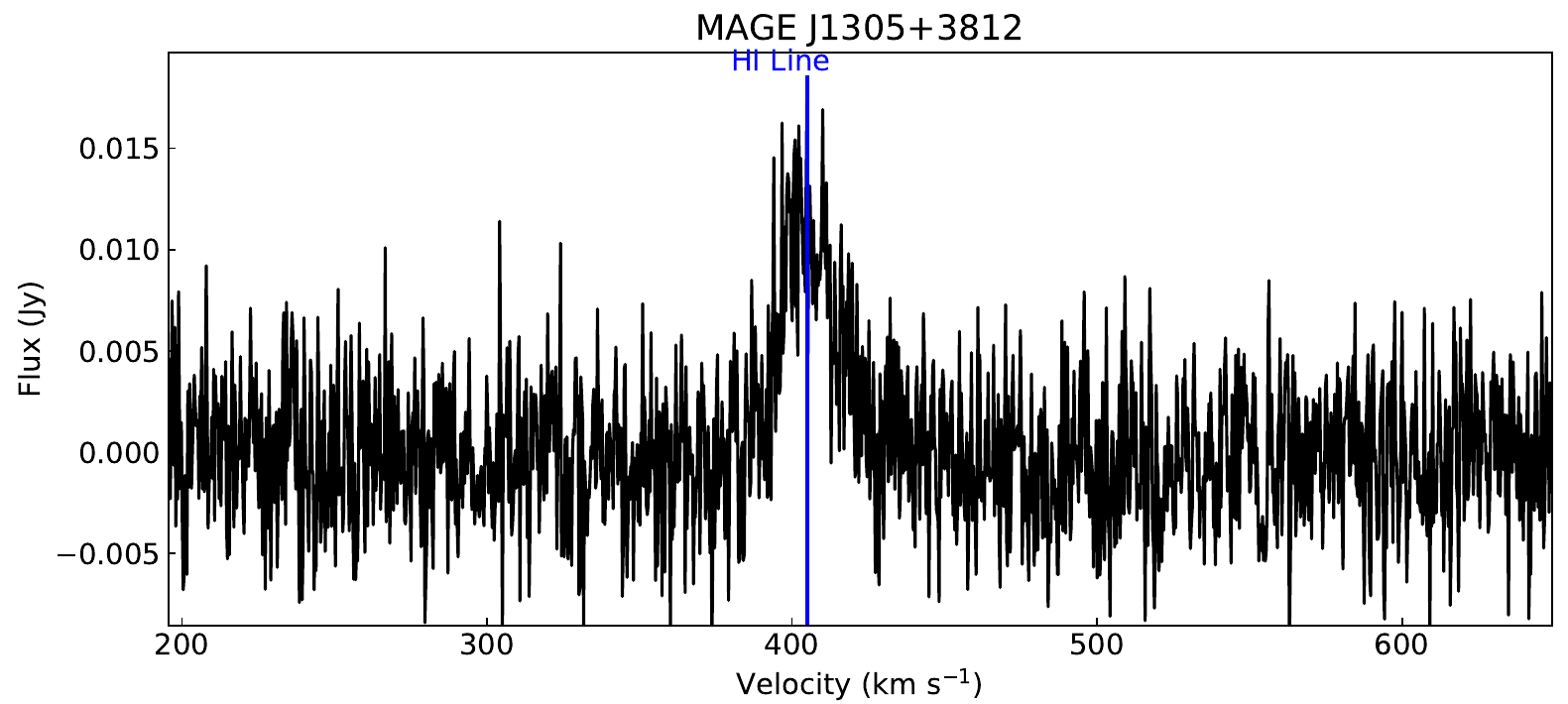}
    \caption{GBT \hi\ spectrum of MAGE J1305+3812 taken as part of project ID GBT24B-319. The \hi\ emission line used to derive the velocity (404$\pm$2\kms) is marked. } 
    \label{1305+3812}
\end{figure*} 

\begin{figure*}
    \centering
    \includegraphics[width=.57\linewidth]{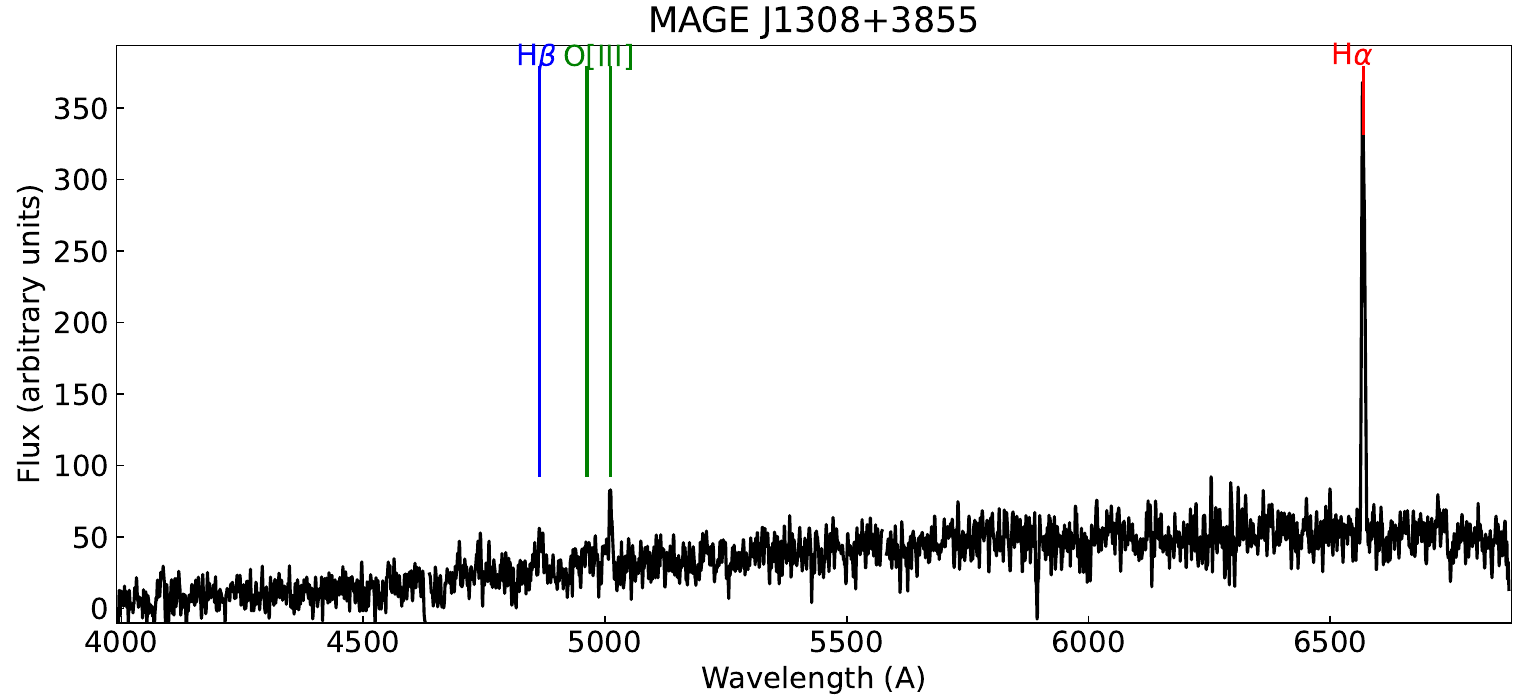}
    \caption{MDM 2.4~m spectrum of MAGE J1308+3855 taken April 2025. The H$\alpha$, H$\beta$, and [\ion{O}{3}] emission lines used to derive the velocity (274$\pm$36\kms) are marked. The spectra are not flux calibrated. } 
    \label{4182_811}
\end{figure*} 

Figures~\ref{0006-2456} to \ref{4182_811} present the observed spectra of the six newly velocity-confirmed candidates.  

\textbf{MAGE J0006-2456} was observed by the GBT as part of project GBT25A-155 (Figure~\ref{0006-2456}). Based on the observed spectrum, we derived a velocity of 570$\pm$2~\kms\ which is in agreement with its host NGC~0024's velocity of 554$\pm$1~\kms. MAGE~J0006-2456 was part of the 2dF Galaxy Redshift Survey, with a velocity of 689$\pm$123\kms, but the uncertainties were too large to confirm it as a satellite \citep{6dFGS}.

\textbf{MAGE J0609-3225} was observed with SALT during  January 2025 with the RSS instrument (Figure~\ref{0609-3225}). Based on the observed spectrum, we derived a velocity of 713$\pm$11~\kms\ which is in agreement with its host ESO364-G29's velocity of 787$\pm$3~\kms.

\textbf{MAGE J1052+3646} was observed in April 2025 with the OSMOS instrument on the MDM 2.4~m telescope (Figure~\ref{3432_365}).  Based on the observed spectrum, we derived a velocity of 523$\pm$40~\kms\ which is in agreement with its host NGC~3432's velocity of 616$\pm$4~\kms. MAGE J1052+3646 is part of DESI which reports a velocity of 570$\pm$10~\kms.

\textbf{MAGE J1052+3628} is part of DESI which derived a velocity of 481$\pm$10~\kms. It was observed by the GBT as part of project GBT25A-155 (Figure~\ref{1052+3628}), however its \hi\ spectrum is contaminated by its host NGC~3432 and another satellite, MAGE~J1052+3635. The detection is 4.6~Jy~\kms\ (\hi\ mass of 8.6$\times 10^7$~M$_\odot$ at the distance of NGC~3432) and is shown in Figure~\ref{1052+3628}. It would have to account for $<$10\% of the observed \hi\ flux to have an \hi mass less than the stellar mass of 7.9$\times 10^6$ M$_\odot$ we derive in \citetalias{idmage}, implying MAGE~J1052+3628 is gas-rich. 

\textbf{MAGE J1305+3812} was observed by the GBT as part of project GBT24B-319 (Figure~\ref{1305+3812}). Based on the observed spectrum, we derived a velocity of 404$\pm$2~\kms\ which is in agreement with its host IC~4182's velocity of 321$\pm$4~\kms.

\textbf{MAGE J1308+3855} was observed in April 2025 with the OSMOS instrument on the MDM 2.4~m telescope (Figure~\ref{4182_811}).  Based on the observed spectrum, we derived a velocity of 274$\pm$36~\kms\ which is in agreement with its host IC~4182's velocity of 321$\pm$4~\kms.  MAGE J1308+3855 is part of DESI which derived a velocity of 270$\pm$10~\kms.  

\section{Background Galaxies and GBT Non-detections}\label{background}

\begin{deluxetable*}{l r r c r r l l r r }
\centerwidetable
\tabletypesize{\footnotesize}
\tablecaption{Velocities of Background Galaxies \label{table:Background_galaxies}}
\tablehead{
\colhead{Name} & \colhead{RA}  & \colhead{Dec} & \colhead{Host}  & \colhead{Host Vel.}  & \colhead{Sat Vel.} & \colhead{Telescope} & \colhead{Exp. Time} &  \colhead{\hi-Flux} & \colhead{Sample} \\
\colhead{}  & \colhead{J2000}  & \colhead{J2000}  & \colhead{ }  & \colhead{\kms} & \colhead{\kms} & \colhead{} & \colhead{sec.} &  \colhead{Jy \kms} & \colhead{} \\
\colhead{(1)} & \colhead{(2)} & \colhead{(3)} & \colhead{(4)} & \colhead{(5)} & \colhead{(6)} & \colhead{(7)} & \colhead{(8)} & \colhead{(9)} & \colhead{(10)}
}
\startdata
MAGE~J1417+5625 & 214.4019 & 56.4252 & NGC~5585 & 303$\pm$1 & 1990$\pm$2 & GBT & 3600 & 0.58$\pm$0.10 & H \\
MAGE~J1418+5559 & 214.7138 & 55.9897 & NGC~5585 & 303$\pm$1 & 6098$\pm$42 & MDM & 3600 & & E \\
MAGE~J1423+5613 & 215.8692 & 56.2323 & NGC~5585 & 303$\pm$1 & 1817$\pm$36 & MDM & 3600 & & H \\
MAGE~J1423+5613 & 215.8692 & 56.2323 & NGC~5585 & 303$\pm$1 & 1806$\pm$2 & GBT & 8400 & 0.25$\pm$0.04 & H \\
\hline 
\enddata
\tablecomments{
Column~1: ID-MAGE identifier Column~2: the Right Ascension (J2000.0). Column~3: the Declination (J2000.0). Column~4: Host name. Column~5: Recessional velocity of the host in \kms. Column~6: Recessional velocity of the candidate in \kms. Column~7: Sample candidate is part of: H is High Likelihood and E is \textbf{Extended Sample} in \citetalias{idmage}. Column~8: Telescope satellite was observed by. Column~9: Exposure time in seconds for GBT, MDM, and SALT observations. Column~10: \hi-Flux for GBT targets.\\
\textbf{See electronic version for the full table.}}
\end{deluxetable*}

\begin{deluxetable*}{l r r r c r c c r r}
\centerwidetable
\tabletypesize{\footnotesize}
\tablecaption{GBT \hi\ Limits for Undetected Galaxies \label{table:GBT_limits}}
\tablehead{
\colhead{Name} & \colhead{RA}  & \colhead{Dec}  & \colhead{Host}  & \colhead{Host Dist}  & \colhead{Exp}  & \colhead{$\sigma_{12}$}  & \colhead{Flux Limit}  & \colhead{S} & \colhead{Vel.} \\
\colhead{}  & \colhead{J2000}  & \colhead{J2000}  & \colhead{ }  & \colhead{Mpc} & \colhead{min}  & \colhead{mJy}  & \colhead{Jy \kms} & \colhead{} & \colhead{\kms} \\
\colhead{(1)} & \colhead{(2)} & \colhead{(3)} & \colhead{(4)} & \colhead{(5)} & \colhead{(6)} & \colhead{(7)} & \colhead{(8)} & \colhead{(9)} & \colhead{10}
}
\startdata
\hline
MAGE~J1208+6934 & 182.0476 & 69.5671 & NGC~4236 & 4.31 & 40 & 2.6 & 0.221 & E & 2280$\pm$10 \\
MAGE~J1209+3853 & 182.4103 & 38.8956 & NGC~4244 & 4.2 & 40 & 0.9 & 0.072 & E \\
MAGE~J1210+3850 & 182.6051 & 38.8340 & NGC~4244 & 4.2 & 20 & 1.7 & 0.144 & H \\
MAGE~J1211+3634 & 182.7909 & 36.5763 & NGC~4244 & 4.2 & 20 & 1.8 & 0.150 & H \\
MAGE~J1211+6939 & 182.8089 & 69.6523 & NGC~4236 & 4.31 & 40 & 1.1 & 0.096 & E \\
MAGE~J1213+3848 & 183.2671 & 38.8160 & NGC~4244 & 4.2 & 60 & 1.1 & 0.090 & H \\
MAGE~J1213+3736 & 183.4591 & 37.6109 & NGC~4244 & 4.2 & 120 & 0.5 & 0.040 & E \\
MAGE~J1214+6903 & 183.5972 & 69.0526 & NGC~4236 & 4.31 & 80 & 1.3 & 0.110 & E  & 2240$\pm$10 \\
\hline 
\enddata
\tablecomments{
Column~1: ID-MAGE identifier Column~2: the Right Ascension (J2000.0). Column~3: the Declination (J2000.0). Column~4: Host name. Column~5: Host distance. Column~6: Exposure time in seconds. Column~7: RMS noise of the GBT data smoothed to 12~\kms\ velocity resolution. Column~8: 5$\sigma$ \hi-flux detection limit.  Column~9: The sample from \citetalias{idmage} the candidate is part of: \textbf{C is Confirmed}, H is High Likelihood, and E is \textbf{Extended Sample}. \textbf{Column~10: Velocities for galaxies undetected by the GBT that are in Table~\ref{table:Background_galaxies}.} \\
\textbf{Galaxies with velocities outside the observed GBT bandpass are not included in this table. } \\
\textbf{See electronic version for the full table.}}
\end{deluxetable*}

Table~\ref{table:Background_galaxies} presents the velocities for candidates which we identify as background galaxies.  For each galaxy, the table includes the presumed host and host's velocity, the candidate's velocity, the telescope it was observed with, the exposure time of the observations, the \hi\ flux, if it was detected with the GBT, and if the candidate was part of the High-Likelihood Sample (labeled H) or the Extended Sample (labeled E). 

Table ~\ref{table:GBT_limits} presents the detection limits for the undetected {satellites and} satellite candidates observed with the GBT.  The detection limits vary from target to target based on the variable exposure times used. For each galaxy, the table includes the presumed host and host's distance, the exposure time in minutes, $\sigma_{12}$, the 5$\sigma$ detection limit and if the candidate was part of the {Confirmed Satellite Sample (labeled C),} High-Likelihood Sample (labeled H), or the Extended Sample (labeled E). {}  

To derive the \hi\ detection limits, we smoothed the data to a velocity resolution of 12~\kms\ and measured the rms noise ($\sigma_{12}$) in the smoothed data. We calculate the $5\sigma$ flux limits as: 
\begin{equation}
     F_{HI}^{\rm lim} = 5 \times 12 \times \sqrt{2} \times \sigma_{12}
\end{equation}
where 12~\kms\ is the channel width, and $\sqrt{2}$ accounts for the two channels over which we expect a signal to be detected assuming an expected \hi\ line width of $\simeq$25~\kms.  {Of the \gbtnodetect\ non-detections, we derive \hi-flux limits for two confirmed satellites, four background galaxies, 27 High-Likelihood Candidates, and 18 Extended Sample Candidates. For MAGE~J1052+3646, we cannot place \hi\ flux limits due to contamination from NGC~3432. For galaxies with velocities in Table~\ref{table:Background_galaxies} outside the observed GBT bandpass, we do not derive \hi\ flux limits as the spectra do not cover the \hi\ flux from the galaxies.} 

We do not derive \hi\ mass limits for the undetected satellite candidates because their distances are unconstrained. The majority of undetected candidates are likely background galaxies outside of the observed GBT bandpass. This is especially likely for the 18 undetected Extended Sample candidates. Our GBT observations cover \hi\ velocities from $\sim$-1000 to 3000 \kms, and, as shown in Appendix Table~\ref{table:Background_galaxies}, about half the contaminants in the High-Likelihood Sample and most in the Extended Sample have velocities above 3000~\kms. Additionally, candidates can be non-detections if its velocity overlaps with MW \hi-emission, which obscures extragalactic \hi\ sources within its velocity range.  Depending on pointing, the unusable velocity range is approximately $-150$ and 150~\kms.  This overlaps the expected velocity range for satellites of ten hosts with $v_{sys}\leq$300~\kms\ (Table~\ref{table:hosts}). NGC~4236, NGC~6503, and UGC~08201 with $v_{sys}\leq$~50~\kms\ are the most impacted. 

\section{MDM and GBT Observations of NGC~3738 Candidates}

\begin{deluxetable*}{l r r c r r r}
\centerwidetable
\tabletypesize{\footnotesize}
\tablecaption{Velocities of Galaxies Near NGC~3738\label{table:Background_3738}}
\tablehead{
\colhead{Name} & \colhead{RA}  & \colhead{Dec} & \colhead{Vel.}  & \colhead{Telescope} & \colhead{Exp. Time} & \colhead{\hi-flux} \\
\colhead{}  & \colhead{J2000}  & \colhead{J2000} & \colhead{\kms} & \colhead{} & \colhead{sec. } &  \colhead{Jy \kms} \\
\colhead{(1)} & \colhead{(2)} & \colhead{(3)} & \colhead{(4)} & \colhead{(5)} & \colhead{(6)} & \colhead{(7)}
}
\startdata
MAGE~J1137+5358 & 174.2724 & 53.9747 & 2801$\pm$38 & MDM & 3600 & \\
MAGE~J1137+5358 & 174.2724 & 53.9747 & 2785$\pm$2 & GBT & 4200 & 0.26$\pm$0.04 \\
MAGE~J1143+5500 & 175.9203 & 55.0011 & 5897$\pm$35 & MDM & 2400 &  \\
\hline 
\enddata
\tablecomments{
Column~1: ID-MAGE identifier Column~2: the Right Ascension (J2000.0). Column~3: the Declination (J2000.0). Column~4: Recessional velocity of the galaxies in \kms. Column~5: Telescope galaxy was observed with  Column~6: Exposure time in seconds for GBT or MDM. Column~7: \hi-flux from the GBT }
\end{deluxetable*}

Table~\ref{table:Background_3738} includes the velocity and \hi-flux measurements for the candidates followed-up around NGC~3738.  With the removal of NGC~3738 from the sample, these galaxies are separated from the other background galaxies to avoid confusion.

\section{Interesting Background Dwarf Galaxies Found in ID-MAGE}\label{LV_dwarfs}

\begin{figure*}
    \centering
    \includegraphics[width=.65\linewidth]{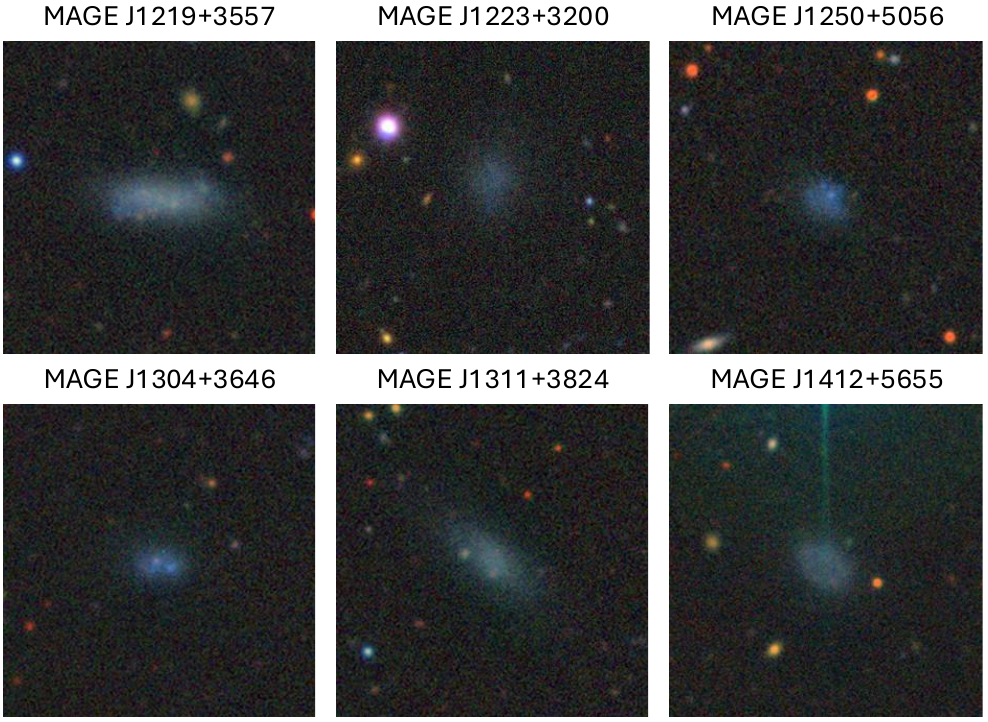}
    \caption{ Cutout color-composite images of six galaxies identified in the ID-MAGE search from the Legacy Surveys DR10. Cutouts are 1.25\arcmin\ per side.  These six galaxies are unassociated with their presumed hosts and may be within the Local Volume based on their velocities in Section~\ref{observations}. } 
    \label{LV_dwarfs_cutouts}
\end{figure*} 

Among the candidates identified as background galaxies based on their velocities is a set of six which may be isolated galaxies within the Local Volume (Figure~\ref{LV_dwarfs_cutouts}).  These galaxies have velocities ranging from 600 to 1000~\kms\ and {$\Delta v_{r} >150$~\kms} compared to their presumed hosts. Based on their surrounding environments, there are indications that a few of them may be isolated, low-mass galaxies. In order to place these within the Local Volume, accurate TRGB distances are required.

\textbf{MAGE~J1219+3557} was found near NGC~4244 ($v$$=$ 245$\pm$1~\kms) and has a velocity of 955$\pm$10 \kms. Behind NGC~4244's is a galaxy group at $\simeq$15.8$\pm$0.4~Mpc and $v\simeq1000$~\kms\ \citep{yuan2020}. MAGE~J1219+3557 may not be a group member as the nearest members ---MCG$+$06$-$27$-$038 and UGC~07257--- have projected distances $\simeq$270~kpc. Thus, it is either on the group outskirts or an isolated galaxy at $\simeq12$~Mpc from its velocity. Assuming a distance of 12~Mpc, it has a stellar mass of $\log(M_*/M_\odot) \simeq$7.2, similar to WLM \citep{Cook14}.  

\textbf{MAGE~J1223+3200} was found around NGC~4395 ($v$$=$ 319$\pm$1~\kms) and has a velocity of 685$\pm$2 \kms.  Based on its velocity and position on the sky, it may be a satellite of NGC~4414 (17.2$\pm$0.88~Mpc \citealt{Bono2010}; 708$\pm$1~\kms\ \citealt{springbob05}).  It lies within the projected virial radius of the MW-analog NGC~4414, and their relative velocity is $<$40~\kms.

\textbf{MAGE~J1250+5056} was found near NGC~4707 ($v$$=$468$\pm$1~\kms) and has a velocity of 694$\pm$2 \kms. NGC~4707 and its satellite UGC~7950 are the closest galaxies to MAGE~J1250+5056 in velocity space. As NGC~4707 is an SMC-mass host, MAGE~J1250+5056 is likely not bound to it.  Based on its velocity, MAGE~J1250+5056 has a distance of $\sim9-10$~Mpc, $M_*\simeq2\times10^6~\msun$, and $M_{HI}\simeq2\times10^6~\msun$. This would make MAGE~J1250+5056 similar to Pavo and Corvus~A in terms of stellar mass \citep{BMP25}. It has an SBF distance of 7.8$^{+1.88,4.39}_{-1.31,2.27}$~Mpc which is in agreement with NGC~4707 at 6.38$\pm$0.29~Mpc, but its relative velocity to NGC~4707 is above our velocity threshold especially as NGC~4707 is an SMC-mass host so we do not consider it a confirmed satellite.

\textbf{MAGE~J1304+3646} ($v$$=$688$\pm$10~\kms) and \textbf{MAGE~J1311+3824} ($v$$=$746$\pm$10~\kms) were found around IC~4182 ($v$$=$ 321~\kms), placing them in close on-sky proximity to MW analog M~94 ($v$$=$ 308~\kms: \citealt{springbob05}). % as IC~4182 is on the outskirts of the M~94 group. 
Both are more than 300~kpc from M~94 and have {$\Delta v_{r} > $300 \kms} compared to M~94. Behind the M~94 group lies the NGC~5005 group. The NGC~5005 group is at $\simeq16$~Mpc and has velocities of 900-1200~\kms\ \citep{Kourkchi17}.  At a distance of 16~Mpc, both fall outside the assumed 300~kpc virial radius for a MW analog but may be part of the larger group.  A third possibility is that they are isolated dwarf galaxies between the two larger galaxy groups. 

\textbf{MAGE~J1412+5655}, found around NGC~5585 ($v$$=$303$\pm$1~\kms), has a velocity of 699$\pm$2~\kms\ and was previously identified as SMDGJ1412582+565531 in the Systematically Measuring Ultra Diffuse Galaxies (SMUDGes; \citealt{SMUDGESV}) survey. There are no known galaxies with a stellar mass of $>10^8$~M$_\odot$, a velocity between 400--1000 \kms, and within 2 degrees on sky of MAGE~J1412+5655.  The closest in projection is NGC~5585 which we rule out as its host.  Based on velocity, we estimate a distance of 9-10~Mpc and place it 2-3~Mpc behind NGC~5585. The lack of semi-resolved features in the Legacy Survey imaging of MAGE~J1412+5655 further supports a distance of $>$6.5~Mpc.  Assuming a distance of 10~Mpc gives a stellar mass of $\log(M_*/M_\odot)\simeq$6.6 and \hi\ mass of $\log(M_{HI}/M_{\odot}$)$\simeq$6.5.  This makes it similar to the newly discovered galaxy Kamino \citep{BMP25}.

\bibliography{ref}{}
\bibliographystyle{aasjournalv7}

\end{document}